\documentclass[amsmath,amssymb,nofootinbib,superscriptaddress]{revtex4-1}

\usepackage{aas_macros,esint,graphicx}
\usepackage[allcolors=blue,colorlinks=true]{hyperref}

\usepackage[compact]{titlesec}         
\titlespacing{\subsection}{10pt}{10pt}{10pt}
\titlespacing{\section}{10pt}{10pt}{10pt}
\numberwithin{equation}{section}

\makeatletter
\renewcommand{\p@subsection}{}
\renewcommand{\p@subsubsection}{}
\makeatother

\newcommand{\ab}[1]{\left|#1\right|}
\newcommand{\av}[1]{\left\langle#1\right\rangle}
\newcommand{\br}[1]{\left[#1\right]}
\newcommand{\cu}[1]{\left\{#1\right\}}
\newcommand{\pa}[1]{\left(#1\right)}

\newcommand{\dt}{\mathop{}\!\delta}
\newcommand{\ed}{\mathop{}\!\mathrm{d}}
\newcommand{\pd}{\mathop{}\!\partial}

\DeclareMathOperator\arctanh{arctanh}
\DeclareMathOperator\im{Im}
\DeclareMathOperator\re{Re}
\DeclareMathOperator\sign{sign}

\def\slr{\mathsf{SL}(2,\mathbb{R})}
\def\tr{\tilde{r}}

\begin{document}

\title{\texorpdfstring{\phantom{}\vspace{90pt}\\
\huge Photon Rings Around Warped Black Holes\vspace{30pt}}
{Photon Rings Around Warped Black Holes}}

\author{\Large Daniel Kapec}
\email{danielkapec@fas.harvard.edu}
\affiliation{\footnotesize Center of Mathematical Sciences and Applications, Harvard University, Cambridge, Massachusetts 02138, USA}
\affiliation{\footnotesize Center for the Fundamental Laws of Nature, Harvard University, Cambridge, Massachusetts 02138, USA}

\author{\Large Alexandru Lupsasca}
\email{alexandru.v.lupsasca@vanderbilt.edu}
\affiliation{\footnotesize Department of Physics and Astronomy, Vanderbilt University, Nashville, Tennessee 37212, USA}

\author{\Large Andrew Strominger}
\email{strominger@physics.harvard.edu}
\affiliation{\footnotesize Center for the Fundamental Laws of Nature, Harvard University, Cambridge, Massachusetts 02138, USA}

\begin{abstract}
\vspace{30pt}
The black hole photon ring is a prime target for upcoming space-based VLBI missions seeking to image the fine structure of astrophysical black holes.
The classical Lyapunov exponents of the corresponding nearly bound null geodesics control the quasinormal ringing of a perturbed black hole as it settles back down to equilibrium, and they admit a holographic interpretation in terms of quantum Ruelle resonances of the microstate dual to the Kerr black hole.
Recent work has identified a number of emergent symmetries related to the intricate self-similar structure of the photon ring.
Here, we explore this web of interrelated phenomena in an exactly soluble example that arises as an approximation to the near-extremal Kerr black hole.
The self-dual warped AdS$_3$ geometry has a photon ring as well as $\slr$ isometries and an exactly calculable quasinormal mode (QNM) spectrum.
We show explicitly that the geometric optics approximation reproduces the eikonal limit of the exact QNM spectrum, as well as the approximate ``near-ring'' wavefunctions.
The $\slr$ isometries are directly related to the emergent conformal symmetry of the photon ring in black hole images but are distinct from a recently discussed conformal symmetry of the eikonal QNM spectrum.
The equivalence of the classical QNM spectrum---and thus the photon ring---to the quantum Ruelle resonances in the context of a spacetime with a putative holographic dual suggests that the photon ring of a warped black hole is indeed part of the black hole hologram.
\end{abstract}

\maketitle

\setcounter{tocdepth}{2}
\tableofcontents

\section{Introduction}

The photon ring around a black hole controls a surprising number of observational signatures relevant to both current and future gravitational experiments.
Its intricate, nested subring structure associated to nearly bound photons predicts distinct interferometric signatures in black hole images \cite{Johnson2020}, and may become visible with near-term space-based missions.
The ring also controls quasinormal ringdowns predicted for LIGO \cite{Abbott2016}: the last tones emitted as a perturbed black hole settles back down to equilibrium consist of nearly trapped gravitational waves slowly leaking off of the photon ring.\footnote{This striking connection is clearly visible in the final frames of \href{https://www.youtube.com/watch?v=I_88S8DWbcU}{this LIGO video} of a simulated collision between two black holes.
The late-time ringdown of the post-merger black hole produces gravitational waves that visibly ripple and decay along the photon ring.}
These two phenomena are closely related in the geometric optics approximation, in which high-frequency solutions of the wave equation are constructed from null geodesic congruences.

Given that the photon ring controls so many distinctive features of classical black hole physics, it is natural to wonder if it might similarly constrain quantum aspects of black holes.
Semi-classical analysis suggests that black holes behave much like isolated quantum-mechanical systems with $e^{S_{\rm BH}}$ approximately thermal microstates and extremely chaotic dynamics.
Known examples of the holographic principle relate the classical quasinormal ringing of the black hole to the damped oscillations of a perturbed thermal quantum state, also known as \textit{Ruelle resonances}.
The photon ring therefore controls, via geometric optics,  the high-frequency sector of the Ruelle spectrum for asymptotically flat black holes.
The quantum-mechanical description of black holes is intimately tied to the emergence of spacetime, and one might think of the photon ring as an emergent geometrization of the high-frequency Ruelle spectrum.
In other words, one can ``hear'' the shape of the photon ring in the spectrum of quantum Ruelle resonances.

\begin{figure}[t]
    \includegraphics[width=\textwidth]{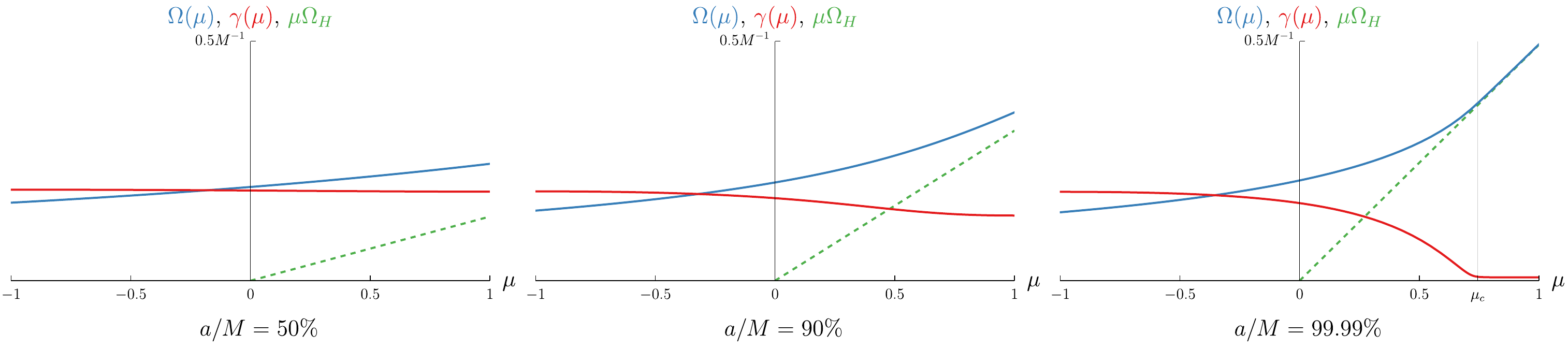}
    \caption{The dispersion relation for a near-extremal black hole  becomes linear for $\mu>\mu_c$ and the Lyapunov exponent vanishes.}
    \label{fig:Dispersion}
\end{figure}

Black hole spectroscopy is quickly becoming an experimental discipline, but no proposed holographic dual to the Kerr black hole has  managed to explain its spectrum of Ruelle resonances from quantum mechanics.
This remains an open challenge for theorists.
The eikonal limit of the Kerr QNM spectrum is \cite{Hadar2022}
\begin{align}
    \omega_{\ell mn}\stackrel{\ell\gg1}{\approx}\pa{\ell+\frac{1}{2}}\Omega(\mu)-i\pa{n+\frac{1}{2}}\gamma_L(\mu)\;,
\end{align}
where $\ell$ is an integer related to the total angular momentum, $-\ell\le m\le\ell$ is the azimuthal angular momentum, and $\mu=m/\ell\in(-1,1)$ is kept fixed as both $\ell$ and $\ab{m}$ are taken large.
The quantity $\mu\approx\sin{\theta_{\rm max}(\tr)}$ determines both the orbital inclination $\theta_{\rm max}$ and the photon shell radius $\tr$ of the bound orbit, and $\Omega(\mu(\tr))$ is the angular velocity of the orbit at radius $\tr$.
The imaginary parts of the QNM frequencies are controlled by $\gamma_L(\mu(\tr))=\gamma(\tr)/\tau(\tr)$, which is the ratio of the bound orbit's Lyapunov exponent $\gamma$ to the orbital period $\tau$ of its libration in $\theta$. 
In Fig.~\ref{fig:Dispersion}, we plot the eikonal dispersion relation and Ruelle exponents for several values of the black hole spin.
For small spins, both the dispersion relation and the Ruelle spectrum are roughly flat, since the spin can be viewed as a small perturbation of the spherically symmetric ($m$-independent) case.
As the spin increases, non-linearities gradually become more pronounced, culminating in the striking graph for the near-extremal black hole.
Reproducing even the qualitative non-linear features of this curve is a sharp diagnostic for any proposed dual to Kerr.

The rightmost plot has an interesting geometric interpretation that will be important for this paper.
As a Kerr black hole spins up and approaches the limiting extremal geometry, a warped throat of divergent proper depth opens up right outside of its event horizon.
This Near-Horizon Extreme Kerr (NHEK) region of \textit{spacetime} exhibits a geometrically realized emergent conformal symmetry that we will denote $\slr_{\rm ISO}$.
The NHEK  geometry is a warped version of the Anti-de Sitter (AdS) throats that emerge outside near-extremal non-rotating black holes.
As the black hole tends to extremality, a subset of the co-rotating bound photon orbits with $\mu>\mu_c$ move into the NHEK region.
Since these geodesics co-rotate with the horizon, $\Omega(\mu)\approx\mu\Omega_H$ is linear in the NHEK regime.
The infinite amount of space in the throat allows the corresponding waves to bounce around for a long time before falling through the horizon or escaping to infinity, so the imaginary parts of the frequencies also go to zero linearly in the deviation from extremality $0\le\kappa=\sqrt{M^2-a^2}\ll1$, or equivalently, in the limit of vanishing Hawking temperature $T_H\approx\kappa/4\pi M$:
\begin{align}
   \omega\approx m\Omega_H-i\pa{n+\frac{1}{2}}2\pi T_H\;.
\end{align}
In other words, the linear ramp and vanishing exponents in the right panel of Fig.~\ref{fig:Dispersion} are essentially a spectral signature of the extremal throat.
The overtone structure with $\gamma_L=\gamma/\tau$ independent of $m$ signals the presence of an exact conformal symmetry geometrically realized in the spacetime as an isometry group.
In the near-extremal throat, the fundamental $n=0$ mode is highest-weight and higher overtones simply correspond to $\slr_{\rm ISO}$-descendants, with the overtone number equal to the descendant level.

The geometric conformal symmetry of the NHEK spectrum does not extend to the non-extremal case or to the portion of the photon ring that remains outside the throat in the extremal limit.
Simply put, the imaginary parts of the frequencies depend on $m$, so no single set of $\slr$ generators $\hat{L}_{\pm}$ could construct overtone families for all of the different values of $m$: the spacings between successive overtones would simply not agree.
However, in a superselection sector of fixed $m$ (and therefore of fixed large real frequency in the geometric optics limit), the overtone structure persists and begs for interpretation.
Schematically, one anticipates an emergent $\slr$ structure in \textit{phase space} rather than in \textit{spacetime}.
Precisely such a structure was identified in \cite{Hadar2022}.
In an appropriately defined ``near-ring region'' of phase space, an emergent conformal symmetry $\slr_{\rm QN}$ controls the eikonal limit of the QNM spectrum.
These symmetry generators are fibered over the photon ring, and their form depends explicitly on the parameters $\gamma$ and $\tau$ associated to each orbit.
Each overtone family decomposes into two highest-weight representations of $\slr_{\rm QN}$ according to the parity of the overtone number, and together the symmetries generate the entire eikonal QNM spectrum.\footnote{In the extremal limit, these combine into a single highest-weight representation of the emergent $\slr_{\rm ISO}$ isometry.}
We are used to thinking of holography in spacetime, but the most general statement of the holographic principle might instead involve phase space.
Does $\slr_{\rm QN}$ have a role to play? We would love to know the answer!

The near-ring region of Kerr exhibits yet another type of emergent conformal symmetry \cite{Hadar2022} that we denote $\slr_{\rm PR}$ and which is singled out for its direct observational relevance: it serves as an organizing principle for the fine structure of photon rings in black hole images.
The Hamiltonian flow generated by the dilations of $\slr_{\rm PR}$ drives unbound geodesics towards the near-ring region of phase space and pushes points on the black hole image closer to the image of the photon ring.
A discrete subgroup relates successive strongly lensed images of the same source, revealing the intricate self-similar structure of the photon ring. 

Clearly, one wishes to understand the relations between these three different conformal groups.
Their relationships are obscured by the complexity of the Kerr black hole, and the discussion would certainly benefit from a simpler exactly solvable toy model.
This raises the question: what is the simplest holographic system with a photon ring controlling its short-wave Ruelle spectrum?

The BTZ black hole is certainly the simplest black hole with a known holographic description, but unfortunately it does not have a photon orbit at finite radius.\footnote{As we will describe in section~\ref{sec:WarpedAdS3}, there is a sense in which the BTZ ``photon shell'' resides at the conformal boundary of the spacetime.}
Black holes in higher-dimensional AdS spacetimes do admit unstably bound null orbits, but they also possess stably trapped orbits that bounce back-and-forth off of the boundary and introduce a timescale longer than that associated to photon rings, thereby obscuring the role of the latter.
The simplest geometry known to us that admits both a photon ring and a reasonably well understood holographic description is the self-dual warped AdS$_3$ spacetime, which is better known as a constant-$\theta$ slice of the near-NHEK throat.

In this paper, we accordingly study photon rings in self-dual warped AdS$_3$.
In section~\ref{sec:WarpedAdS3}, we begin with a brief review of this spacetime and then show that, like NHEK, it enjoys an $\slr_{\rm ISO}$ isometry group that makes the wave equation exactly solvable in terms of hypergeometric functions.
The fundamental $n=0$ QNMs are highest-weight with respect to this symmetry algebra, and the $n>0$ overtones arise as $\slr_{\rm ISO}$-descendants.
The geometry has a photon orbit at finite radius.
We show explicitly that the corresponding geometric optics approximation reproduces the eikonal limit of the exact QNM spectrum, as well as the approximate near-ring wavefunctions.
Interestingly, we find that in order to recover the entire eikonal QNM spectrum from the geometric optics approximation, it is necessary to also consider a second bound null geodesic located behind the horizon.

Next, we define the $\slr_{\rm QN}$ generators that operate exclusively within the near-ring region of phase space.
Each QNM overtone family decomposes into two highest-weight representations of $\slr_{\rm QN}$, so $\slr_{\rm QN}$ does not reduce to $\slr_{\rm ISO}$ in the extremal limit.
These groups arise from distinct constructions, but their action on the QNM wavefunctions is related in a simple way in the near-ring region, explained below.
It seems plausible that there is a more fundamental relationship between the two symmetries, but we leave this question to future investigation.

In section~\ref{sec:Symmetries}, we define the conformal group $\slr_{\rm PR}$ that acts on the phase space of null geodesics and contains a dilation that scales every light ray into the photon ring.
Interestingly, we find that the observationally relevant $\slr_{\rm PR}$ and the geometric $\slr_{\rm ISO}$ are both special cases of the more general $\slr$ phase space construction explored in \cite{Hadar2022}, and that $\slr_{\rm PR}$ can be thought of as a particular deformation of $\slr_{\rm ISO}$ within this family.

Finally, in section~\ref{sec:Matching}, we discuss how the spectrum of resonances of self-dual warped AdS$_3$ resembles that of a CFT$_2$ with the corresponding left and right temperatures and a peculiar identification of momenta.
Although the interpretation of this match is still not rigorously understood, 
we take this as evidence that the holographic dual to this black hole spacetime encodes its photon ring, or in other words, that the photon ring is part of the hologram.

\section{Warped \texorpdfstring{AdS$_3$}{AdS3} black holes}
\label{sec:WarpedAdS3}

In this section, we describe a soluble example in the context of a known holographic duality in string theory, in which photon ring Lyapunov exponents and late-time Ruelle resonances can be shown to agree.
The example is self-dual warped AdS$_3$, which is also the near-horizon region of the near-extreme warped AdS$_3$ black hole.
Warped AdS$_3$ is a spacetime in which the $\slr_R\times\slr_L$ isometry of AdS$_3$ is broken down to $\slr_R\times\mathsf{U}(1)$, e.g., by fluxes or a gravitational Chern-Simons term.
It is characterized by a continuous warp factor $\Lambda$ defined such that $\Lambda=1$ for unwarped AdS$_3$.
Such geometries arise in string theory, including examples with exactly soluble string worldsheet CFTs \cite{Israel2005,Detournay2011,Azeyanagi2013}.
On the boundary side, they are related to warped CFTs \cite{Hofman2011,Detournay2012} that are integrable $\bar{J}T$ deformations of ordinary CFTs \cite{Bzowski2019,Guica2018,Chakraborty2018,Apolo2018,Chakraborty2019,Apolo2020,Guica2022}.
Warped AdS$_3$ black holes are also of special interest due to their many similarities to Kerr black holes.
In fact, in the near-extreme near-horizon limit, three-dimensional sections of Kerr with fixed polar angle $\theta$ reduce to a self-dual warped AdS$_3$ quotient with warp factor $\Lambda(\theta)=\frac{\sin{\theta}}{1+\cos^2{\theta}}$.
The literature on these topics is extensive, including, e.g., \cite{Anninos2009,Chen2009,Chen2010a,Chen2010b,Song2012}.

For our present purposes, we are interested in warped AdS$_3$ with $\Lambda>1$ both because it possesses a photon shell and because this is the geometry that describes  polar cross sections of Kerr.
Here, we show that the Lyapunov exponent that governs the exponential deviation of nearly bound orbits from their orbital radius in the photon shell also controls the imaginary part of QNM frequencies in the eikonal limit.
This imaginary part can also be identified with $2\pi T_H$ (with $T_H$ the Hawking temperature of the black hole) times the (real part of the) weight of the associated operator in the dual warped CFT$_2$, as expected on general grounds.
This number is identified under bulk-boundary duality as the Ruelle exponent governing the return to equilibrium of a slightly perturbed thermal quantum state.

The results of this section show, in this toy stringy example,  that the photon shell of a black hole describes properties of its quantum microstates in a thermal ensemble.

\subsection{The near-ring region}
\label{subsec:NearRingRegion}

The metric describing the near-horizon region of the near-extreme warped AdS$_3$ black hole can be obtained from the near-NHEK (near-extreme near-horizon Kerr) geometry by considering hypersurfaces of constant polar angle \cite{Moussa2003,Bouchareb2007,Bredberg2010}:
\begin{align}
	\label{eq:WAdS3}
	ds^2=\ell^2\br{-r\pa{r+4\pi T_H}\ed t^2+\frac{\ed r^2}{r\pa{r+4\pi T_H}}+\Lambda^2\br{\ed\phi+\pa{r+2\pi T_H}\ed t}^2}\;.
\end{align}
In these coordinates, the horizon is at $r=0$, $T_H$ denotes the dimensionless Hawking temperature, and $\ell$ is the (warped) AdS$_3$ radius.
The warp factor $\Lambda$ is defined such that \eqref{eq:WAdS3} reduces to (unwarped) self-dual AdS$_3$ when $\Lambda=1$.

We are interested in affinely parameterized null geodesics $x^\mu(s)=(t(s),r(s),\phi(s))$.
These have the conserved energy 
\begin{align}
	\label{eq:Energy}
	\frac{E}{\ell^2}=r\pa{r+4\pi T_H}\dot{t}-\Lambda^2\pa{r+2\pi T_H}\br{\dot{\phi}+\pa{r+2\pi T_H}\dot{t}}
\end{align}
and angular momentum 
\begin{align}
	\frac{L}{\ell^2}=\Lambda^2\br{\dot\phi+\pa{r+2\pi T_H}\dot t}\;,
\end{align}
where the dot denotes $\pd_s$.
The null condition can be written as
\begin{align}
	\frac{r\pa{r+4\pi T_H}}{\ell^2}g_{\mu\nu}\dot{x}^\mu\dot{x}^\nu=\dot r^2+\frac{V(r)}{\ell^4}
	=0\;,
\end{align}
with
\begin{align}
	\label{eq:Potential}
	V(r)=\frac{L^2}{\Lambda^2}r\pa{r+4\pi T_H}-\br{E+\pa{r+2\pi T_H}L}^2\;.
\end{align}
Equations \eqref{eq:Energy}--\eqref{eq:Potential} imply the null geodesic equation for $x^\mu(s)$.
Bound photon orbits are defined by the conditions
\begin{align}
	\label{eq:OrbitalConditions}
	V(r)=V'(r)
	=0\;.
\end{align}
These depend only on the specific (energy-rescaled) angular momentum
\begin{align}
	\lambda=\frac{L}{E}\;.
\end{align}
The conditions \eqref{eq:OrbitalConditions} must be simultaneously solved for the critical impact parameter $\lambda=\tilde{\lambda}$ and the orbital radius $r=\tr$ of the bound geodesic.
Demanding that $V'(\tr)=0$ requires that
\begin{align}
	\tilde{\lambda}(\tr)=-\frac{1}{\pa{\tilde r+2\pi T_H}\pa{1-\frac{1}{\Lambda^2}}}\;.
\end{align}
Substituting this into the constraint $V(\tr)=0$ yields a quadratic equation with roots at the two critical radii
\begin{align}
    \label{eq:CriticalImpact}
	\tr_\pm=2\pi T_H\pa{\pm\frac{1}{\sqrt{1-\frac{1}{\Lambda^2}}}-1}
	\gtrless0
	\qquad\Longrightarrow\qquad
	\tilde{\lambda}_\pm=\mp\frac{1}{2\pi T_H\sqrt{1-\frac{1}{\Lambda^2}}}
	\lessgtr0\;.
\end{align}
This spacetime therefore admits two bound photon orbits with orbital period $\tau$ and angular velocities $\tilde{\Omega}_\pm$ given by
\begin{align}
	\tau=\mp2\pi\left.\frac{\dot t}{\dot\phi}\right|_{r=\tr_\pm,\lambda=\tilde{\lambda}_\pm}
	=\frac{1}{T_H\sqrt{1-\frac{1}{\Lambda^2}}}
	>0,\qquad
	\tilde{\Omega}_\pm=\mp\frac{2\pi}{\tau}
	\lessgtr0\;.
\end{align}
Since $\tr_-<0<\tr_+$, the inner orbit lies inside the horizon.
We will thus restrict our attention to the outer orbit and consider nearly bound null geodesics with small radial deviation $\dt r\equiv r-\tr_+$.
More specifically, we are interested in the near-ring region defined in \textit{phase space} by
\begin{align}
    \label{eq:NearRingRegion}
    \text{NEAR-RING REGION:}\qquad
    \begin{cases}
        \ab{\dt r}\ll T_H&\qquad\text{(near-peak)}\;,\\
        \displaystyle\ab{\lambda-\tilde{\lambda}_+}\ll\frac{1}{T_H}&\qquad\text{(near-critical)}\;,\\
        \displaystyle\frac{\ell^2}{E}\ll\frac{1}{T_H}&\qquad\text{(high-energy)}\;.
    \end{cases}
\end{align}
The first condition zooms in on the bound orbit in spacetime, while the second condition zooms in on the bound orbit in momentum space.
Together, these conditions scale into the (outer) \textit{photon shell}, defined as the phase space locus
\begin{align}
    \label{eq:PhotonShell}
    \text{PHOTON SHELL:}\qquad
    \dt r=0=\lambda-\tilde{\lambda}_+.
\end{align}
The last condition in \eqref{eq:NearRingRegion} is required in order to relate solutions of the wave equation to geodesic congruences and will only become important in subsequent sections.
As $T_H\to0$, we note that the near-ring region completely localizes on the bound orbit in physical space ($\dt r=0$) but also fills all of momentum space (with no restriction on $E$ or $\lambda$).

The radial deviation of a nearly bound orbit grows exponentially as $\dt r\sim e^{\gamma n}$, where $n$ is the orbit number and $\gamma$ is a Lyapunov exponent that can be determined by linearizing the geodesic equation about the near-ring region:
\begin{align}
	\dt\ddot r\approx-\frac{1}{2\ell^4}V''(\tilde r)\dt r\;.
\end{align}
Since $\tr_+$ is a local maximum of the radial potential,
\begin{align}
	V''(\tilde r_+)=-\frac{E^2}{2\pi^2T_H^2}
	<0\;,
\end{align}
the motion is indeed unstable and one finds the exponential deviation
\begin{align}
	\dt r(s)\approx\exp\pa{\frac{E}{2\pi T_H\ell^2}s}\dt r(0)\;.
\end{align}
The lapse in the affine parameter $s$ per orbit is
\begin{align}
	 \Delta s=\int_0^{2\pi}\frac{\ed\phi}{\dot{\phi}}
	 =\left.\frac{2\pi}{\dot{\phi}}\right|_{r=\tr_+,\lambda=\tilde{\lambda}_+}
	 =\frac{(2\pi\ell)^2T_H}{E\sqrt{1-\frac{1}{\Lambda^2}}}\;.
\end{align}
It follows that, as a function of the orbit number $n=\frac{\Delta\phi}{2\pi}$, the radial deviation $\dt r(n)$ increases as
\begin{align}
    \label{eq:GeodesicDeviation}
	\dt r(n)\approx e^{\gamma n}\dt r(0)\;,\qquad
	\gamma=\frac{2\pi}{\sqrt{1-\frac{1}{\Lambda^2}}}\;.
\end{align}
Together with $\phi\approx\tilde{\Omega}_+t$, this completes the solution of the null geodesic equation in the near-ring region \eqref{eq:NearRingRegion} of the warped AdS$_3$ metric \eqref{eq:WAdS3}, where much of the interesting motion in the black hole exterior occurs.
A similar analysis applies near the inner photon shell $r-\tr_-=0=\lambda-\tilde{\lambda}_-$ in the black hole interior, resulting in the same Lyapunov exponent $\gamma$.
The solution everywhere else in the spacetime can be written in terms of the integrals evaluated in \cite{Kapec2020}, but this full solution will not be of interest to us here.

To later connect with the QNM spectrum, we note here the simple relations
\begin{align}
	\label{eq:LyapunovExponent}
	\tilde{\Omega}_\pm=\frac{1}{\tilde{\lambda}_\pm}
	=\mp2\pi T_H\sqrt{1-\frac{1}{\Lambda^2}}\;,\qquad
	\gamma_L\equiv\frac{\gamma}{\tau}
	=2\pi T_H\;.
\end{align}
In the limit $\Lambda\to1$, the quantities $\tr_\pm$, $\gamma$, and $\tau$ diverge, but the ratio $\gamma_L=\frac{\gamma}{\tau}$ remains finite.
In this sense, the unwarped AdS$_3$ photon shell \eqref{eq:PhotonShell} resides at the conformal boundary.

\subsection{Quasinormal mode spectrum}
\label{subsec:Spectrum}

In this section, we compute the QNM frequencies of the (self-dual) warped AdS$_3$ spacetime \eqref{eq:WAdS3}, whose imaginary parts become the Ruelle exponents.
We are primarily interested in photons, but since the effects of spin are subleading in the geometric optics regime, it is sufficient for our purposes to solve the wave equation for a massless scalar field $\Phi(x^\mu)$ in the background \eqref{eq:WAdS3}:
\begin{align}
	\nabla^2\Phi=0\;.
\end{align}
Inserting the ansatz $\Phi(t,r,\phi)=e^{-i\omega t+im\phi}\psi(r)$, this reduces to a second-order ODE for the radial function $\psi(r)$,
\begin{align}
	r\pa{r+4\pi T_H}\psi''(r)+\pa{2r+4\pi T_H}\psi'(r)+\br{\frac{\br{\omega+m\pa{r+2\pi T_H}}^2}{r\pa{r+4\pi T_H}}-m^2-\beta^2+\frac{1}{4}}\psi(r)=0\;,
\end{align}
where we have defined 
\begin{align}
	\label{eq:Beta}
	\beta=im\sqrt{\pa{1-\frac{1}{\Lambda^2}}-\frac{1}{4m^2}}\;.
\end{align}
This is a form of the hypergeometric equation and the solutions are given in terms of $\hat{\omega}=\frac{\omega}{2\pi T_H}$ as
\begin{align}
    \psi^{\rm in}(r)&=r^{-\frac{i}{2}\pa{\hat{\omega}+m}}\pa{\frac{r}{4\pi T_H}+1}^{\frac{i}{2}\pa{\hat{\omega}-m}}{_2F_1}\br{\frac{1}{2}+\beta-im,\frac{1}{2}-\beta-im;1-i\pa{\hat{\omega}+m};-\frac{r}{4\pi T_H}}\;,\\
	\psi^{\rm out}(r)&=r^{+\frac{i}{2}\pa{\hat{\omega}+m}}\pa{\frac{r}{4\pi T_H}+1}^{\frac{i}{2}\pa{\hat{\omega}-m}}{_2F_1}\br{\frac{1}{2}+\beta+i\hat{\omega},\frac{1}{2}-\beta+i\hat{\omega};1+i\pa{\hat{\omega}+m};-\frac{r}{4\pi T_H}}\;.
\end{align}
\normalsize
The first solution $\psi^{\rm in}(r)$ is ``ingoing'' in the sense that a local observer at the horizon will see infalling particles going into the black hole.
The second solution $\psi^{\rm out}(r)$ has particles coming out of the past horizon and therefore does not correspond to QNMs, but rather to ``anti-QNMs'' (App.~\ref{app:AQNM}).
Near the horizon,
\begin{align}
	\psi^{\rm in}(r)\stackrel{r\to0}{\approx}r^{-\frac{i}{2}\pa{\hat{\omega}+m}}\;,\qquad
	\psi^{\rm out}(r)\stackrel{r\to0}{\approx}r^{+\frac{i}{2}\pa{\hat{\omega}+m}}\;,
\end{align}
while for $r\gg4\pi T_H$,
\begin{align}
    \label{eq:BoundaryIngoingModes}
	\psi^{\rm in}(r)&\stackrel{r\to\infty}{\approx}\frac{\Gamma(2\beta)\Gamma\br{1-i\pa{\hat{\omega}+m}}}{\Gamma\pa{\frac{1}{2}+\beta-im}\Gamma\pa{\frac{1}{2}+\beta-i\hat{\omega}}}\pa{4\pi T_H}^{\frac{1}{2}-\beta-\frac{i}{2}\pa{\hat{\omega}+m}}r^{-\frac{1}{2}+\beta}+\pa{\beta\to-\beta}\;,\\
	\label{eq:BoundaryOutgoingModes}
	\psi^{\rm out}(r)&\stackrel{r\to\infty}{\approx}\frac{\Gamma(2\beta)\Gamma\br{1+i\pa{\hat{\omega}+m}}}{\Gamma\pa{\frac{1}{2}+\beta+im}\Gamma\pa{\frac{1}{2}+\beta+i\hat{\omega}}}\pa{4\pi T_H}^{\frac{1}{2}-\beta+\frac{i}{2}\pa{\hat{\omega}+m}}r^{-\frac{1}{2}+\beta}+\pa{\beta\to-\beta}\;.
\end{align}
Resonances occur when one of these terms vanishes.
For the ``in'' modes, the resonant frequencies are of the form
\begin{align}
	\label{eq:ResonantCondition}
	\frac{1}{2}\mp\beta-i\hat{\omega}=-n
\end{align}
for some integer $n\in\mathbb{N}=\cu{0,1,2,\ldots}$.
Inserting \eqref{eq:Beta} and recalling that $\hat{\omega}=\frac{\omega}{2\pi T_H}$, \eqref{eq:ResonantCondition} takes the explicit form
\begin{align}
	\frac{1}{2}\mp im\sqrt{\pa{1-\frac{1}{\Lambda^2}}-\frac{1}{4m^2}}-\frac{i\omega}{2\pi T_H}=-n\;.
\end{align}
Solving for $\omega$ then yields the spectrum of resonant frequencies
\begin{align}
    \label{eq:QNM}
	\omega_{mn\pm}=\mp2\pi T_H m\sqrt{\pa{1-\frac{1}{\Lambda^2}}-\frac{1}{4m^2}}-i\pa{n+\frac{1}{2}}2\pi T_H\;,\qquad
	m\in\mathbb{Z}\;,\quad
	n\in\mathbb{N}\;.
\end{align}
Since $\im\omega_{mn\pm}<0$, the associated modes decay exponentially in the future, so we recognize $\omega_{mn\pm}$ to be the spectrum of QNM frequencies.
Likewise, the anti-QNM frequencies are resonances of the ``out'' modes (App.~\ref{app:AQNM}).

In the eikonal regime $\ab{m}\gg1$, the QNM spectrum \eqref{eq:QNM} becomes
\begin{align}
	\omega_{mn\pm}&\stackrel{\ab{m}\gg1}{\approx}\mp2\pi T_Hm\sqrt{1-\frac{1}{\Lambda^2}}-i\pa{n+\frac{1}{2}}2\pi T_H\\
    \label{eq:EikonalQNMs}
	&=m\tilde{\Omega}_\pm-i\pa{n+\frac{1}{2}}\gamma_L\;,
\end{align}
which shows that the photon shell controls the QNM spectrum $\omega_{mn\pm}$ in the eikonal limit.
More precisely, the outer photon shell governs the branch with $m/\re\omega_{mn+}=\tilde{\lambda}_+<0$, while the inner photon shell governs the branch with $m/\re\omega_{mn-}=\tilde{\lambda}_->0$.
As expected, the full photon shell is needed to recover the complete QNM spectrum.

A closely related calculation in the near-horizon region of (near-)extreme Kerr \cite{Bredberg2010} shows that $h=\frac{1}{2}+\beta$ is the conformal weight of a massless scalar under the $\slr$ isometry of warped AdS$_3$, and hence also of the associated boundary operator in the dual field theory. 
Equation \eqref{eq:ResonantCondition} simply reflects the fact that the thermal two-point function of such a primary field with $\re(h)=\frac{1}{2}$ has the leading long-time behavior $e^{-2\pi T_Ht}$ in the Gibbs state.

In the context of AdS$_3$ holography, BTZ black holes are dual to thermal states in CFT$_2$, and the thermal two-point function takes a universal form that can be recovered by summing over the spectrum of QNMs \cite{Birmingham2002,Birmingham2003}.
Likewise, in the context of warped AdS$_3$ holography, summing over QNMs with frequencies \eqref{eq:EikonalQNMs} produces a hyperbolic sine, which matches the form of the thermal two-point correlator in the putative dual warped CFT$_2$ \cite{Song2018}.

\subsection{Conformal symmetry of the quasinormal mode spectrum}
\label{subsec:ConformalSymmetry}

For the remainder of our discussion, it is convenient to work in the coordinates
\begin{align}
    \label{eq:SimpleCoordinates}
	T=2\pi T_H t\;,\qquad
	x=\log\sqrt{1+\frac{4\pi T_H}{r}}\;,\qquad
	r=\frac{4\pi T_H}{e^{2x}-1}\;.
\end{align}
The horizon is now located at $x\to\infty$ while the boundary resides at $x=0$, and the metric takes the simple form
\begin{align}
	\label{eq:SimpleWAdS3}
	ds^2=\ell^2\br{\frac{-\ed T^2+\ed x^2}{\sinh^2{x}}+\Lambda^2\pa{\ed\phi+\frac{\ed T}{\tanh{x}}}^2}\;.
\end{align}
The isometry group is $\slr\times\mathsf{U}(1)$ with generators  
\begin{align}
    \label{eq:Isometries}
	\hat{L}_0^\mu\pd_\mu=-\pd_T\;,\qquad
	\hat{L}_\pm^\mu\pd_\mu=e^{\pm T}\pa{-\cosh{x}\pd_T\mp\sinh{x}\pd_x+\sinh{x}\pd_\phi}\;,\qquad
	W_0^\mu\pd_\mu=\pd_\phi\;,
\end{align}
which obey the $\slr\times\mathsf{U}(1)$ commutation relations
\begin{align}
    \label{eq:SL2R}
    \br{\hat{L}_0,\hat{L}_\pm}=\mp\hat{L}_\pm\;,\qquad
    \br{\hat{L}_+,\hat{L}_-}=2\hat{L}_0\;,\qquad
    \br{W_0,\hat{L}_m}=0\;.
\end{align}
The Killing vectors define quantities $\hat{L}_m=\hat{L}_m^\mu p_\mu$ that are conserved along geodesics and for our purposes best thought of as Hamiltonian functions in the six-dimensional phase space $(T,x,\phi,p_T,p_x,p_\phi)$ with canonical symplectic form:
\begin{align}
    \label{eq:Functions}
    \hat{L}_0=-p_T\;,\qquad
    \hat{L}_\pm=e^{\pm T}\br{-p_T\cosh{x}+\pa{p_\phi\mp p_x}\sinh{x}}\;,\qquad
    W_0=p_\phi\;.
\end{align}
The action of the spacetime vectors \eqref{eq:Isometries} then lifts in the obvious way to a phase space action, with the Hamiltonian functions \eqref{eq:Functions} acting via the Poisson bracket (with $\cu{x^\mu,p_\mu}=-1$) and satisfying the $\slr\times\mathsf{U}(1)$ algebra \eqref{eq:SL2R}.

The inverse metric is reducible in terms of the Killing vectors
\begin{align}
    \ell^2 g^{\mu\nu}=\hat{L}_0^\mu\hat{L}_0^\nu-\frac{\hat{L}_+^\mu\hat{L}_-^\nu+\hat{L}_+^\nu\hat{L}_-^\mu}{2}-\pa{1-\frac{1}{\Lambda^2}}W_0^\mu W_0^\nu\;,
\end{align}
so the Laplacian may be expressed as the Casimir of $\slr\times\mathsf{U}(1)$:
\begin{align}
	\nabla^2\Phi=-\frac{1}{\ell^2}\br{-\mathcal{L}_{\hat{L}_0}^2+\frac{\mathcal{L}_{\hat{L}_+}\mathcal{L}_{\hat{L}_-}+\mathcal{L}_{\hat{L}_-}\mathcal{L}_{\hat{L}_+}}{2}+\pa{1-\frac{1}{\Lambda^2}}\mathcal{L}_{W_0}^2}\Phi\;.
\end{align}
The mode ansatz $\Phi(T,x,\phi)=e^{-i\hat{\omega}T+im\phi}\psi(x)$ reduces the massless wave equation $\nabla^2\Phi=0$ to the radial ODE
\begin{align}
    \label{eq:RadialODE}
	\psi''(x)+V(x)\psi(x)=0\;,\qquad
	V(x)=\hat{\omega}^2+m^2+2\hat{\omega}m\coth{x}+\frac{\frac{1}{4}-\beta^2}{\sinh^2{x}}\;,
\end{align}
with $\beta$ as in \eqref{eq:Beta}.
We note that $\hat{\omega}T=\omega t$.
The potential satisfies $V(\tilde{x}_\pm)=V'(\tilde{x}_\pm)=0$ when $m=\tilde{\lambda}_\pm\omega=\hat{\lambda}_\pm\hat{\omega}$ with
\begin{align}
	\label{eq:Criticality}
	\hat{\lambda}_\pm=2\pi T_H\tilde{\lambda}_\pm
	=\mp\frac{1}{\sqrt{1-\frac{1}{\Lambda^2}}}\;,\qquad
	\tilde{x}_\pm=\pm\arctanh\sqrt{1-\frac{1}{\Lambda^2}}\;,\qquad
	\hat{\Omega}_\pm=\frac{1}{\hat{\lambda}_\pm}\;.
\end{align}
In particular, one has $\coth{\tilde{x}_\pm}=-\hat{\lambda}_\pm$.
The two independent solutions of the radial ODE \eqref{eq:RadialODE} are
\begin{align}
	\psi^{\rm in}(x)&=e^{i\hat{\omega}x}\pa{\sinh{x}}^{im}{_2F_1}\br{\frac{1}{2}+\beta-im,\frac{1}{2}-\beta-im;1-i\pa{\hat{\omega}+m};\frac{1}{1-e^{2x}}}\;,\\
	\psi^{\rm out}(x)&=e^{-imx}\pa{\sinh{x}}^{-i\hat{\omega}}{_2F_1}\br{\frac{1}{2}+\beta+i\hat{\omega},\frac{1}{2}-\beta+i\hat{\omega};1+i\pa{\hat{\omega}+m};\frac{1}{1-e^{2x}}}\;,
\end{align}
with asymptotics at the boundary $x=0$ given by
\begin{align}
    \label{eq:IngoingModesBoundary}
	\psi^{\rm in}(x)&\stackrel{x\to0}{\approx}\frac{2^{-im}(2x)^{\frac{1}{2}-\beta}\Gamma(2\beta)\Gamma\br{1-i\pa{\hat{\omega}+m}}}{\Gamma\pa{\frac{1}{2}+\beta-im}\Gamma\pa{\frac{1}{2}+\beta-i\hat{\omega}}}+\pa{\beta\to-\beta}\;,\\
    \label{eq:OutgoingModesBoundary}
	\psi^{\rm out}(x)&\stackrel{x\to0}{\approx}\frac{2^{i\hat{\omega}}(2x)^{\frac{1}{2}-\beta}\Gamma(2\beta)\Gamma\br{1+i\pa{\hat{\omega}+m}}}{\Gamma\pa{\frac{1}{2}+\beta+im}\Gamma\pa{\frac{1}{2}+\beta+i\hat{\omega}}}+\pa{\beta\to-\beta}\;.
\end{align}
We have already analyzed the resonances of these modes in section~\ref{subsec:NearRingRegion}.
For the ``in'' modes, resonances arise when $\frac{1}{2}\mp\beta-i\hat{\omega}=-n$, resulting in the spectrum \eqref{eq:QNM} of QNM frequencies.
The QNMs are thus
\begin{align}
    \label{eq:QNMx}
	\Phi_{mn\pm}(T,x,\phi)&=e^{-i\hat{\omega}T+im\phi+i\hat{\omega}x}\pa{\sinh{x}}^{im}{_2F_1}\br{\frac{1}{2}+\beta-im,\frac{1}{2}-\beta-im;1-i\pa{\hat{\omega}+m};\frac{1}{1-e^{2x}}}\;,\\
    \label{eq:FrequenciesQNM}
	\hat{\omega}=\hat{\omega}_{mn\pm}
	&=\pm i\beta-i\pa{n+\frac{1}{2}}\;,
\end{align}
where the two branches $\pm$ of the spectrum are associated with the outer and inner photon shell, respectively.
The anti-QNMs are similarly obtained as resonances of the ``out'' modes in App.~\ref{app:AQNM}.
In the eikonal regime $\ab{m}\gg1$,
\begin{align}
	\label{eq:EikonalBeta}
	\beta\stackrel{\ab{m}\gg1}{\approx}im\sqrt{1-\frac{1}{\Lambda^2}}
	=\mp im\hat{\Omega}_\pm
	\qquad\implies\qquad
	\pm i\beta\stackrel{\ab{m}\gg1}{\approx}m\hat{\Omega}_\pm\;,
\end{align}
so $\hat{\omega}_{mn\pm}\stackrel{\ab{m}\gg1}{\approx}m\hat{\Omega}_\pm-i\pa{n+\frac{1}{2}}$ as expected from \eqref{eq:EikonalQNMs}, and the fundamental $n=0$ mode behaves as
\begin{align}
	\Phi_{m0\pm}(T,x,\phi)&\stackrel{\ab{m}\gg1}{\approx}2^{\frac{1}{2}+im\hat{\Omega}_\pm-im}e^{-\pa{\frac{1}{2}+im\hat{\Omega}_\pm}T+im\phi+imx}\pa{\sinh{x}}^{\frac{1}{2}+im\hat{\Omega}_\pm}\;.
\end{align}
From now on, we focus on the outer bound orbit at $\tilde{x}=\tilde{x}_+$ with angular velocity $\hat{\Omega}=\hat{\Omega}_+$, which controls the $+$ branch $\omega_{mn}=\omega_{mn+}$ of the QNM spectrum.
Suppressing the $+$  subscript and dropping constant factors, we obtain
\begin{align}
	\label{eq:EikonalQNM}
	\Phi_{m0}(T,x,\phi)&\stackrel{\ab{m}\gg1}{\sim}e^{-\pa{\frac{1}{2}+im\hat{\Omega}}T+im\phi+imx}\pa{\sinh{x}}^{\frac{1}{2}+ im\hat{\Omega}}\;.
\end{align}

Independent of the eikonal approximation, one can verify that the $n=0$ fundamental QNMs are highest-weight and that their $n>0$ overtones are $\slr$-descendants.
Indeed, the modes
\begin{align}
    \label{eq:HWQNM}
	\Phi_{m0}^{\rm HW}(T,x,\phi)&=e^{-\hat{h}T+im\phi+imx}\pa{\sinh{x}}^{\hat{h}}\;,\qquad
	\hat{h}=\frac{1}{2}-\beta
	\stackrel{\ab{m}\gg1}{\approx}\frac{1}{2}+im\hat{\Omega}\;,
\end{align}
satisfy
\begin{align}
	\mathcal{L}_{\hat{L}_+}\Phi_{m0}^{\rm HW}=0\;,\qquad
	\mathcal{L}_{\hat{L}_0}\Phi_{m0}^{\rm HW}=\hat{h}\Phi_{m0}^{\rm HW}\;,\qquad
	\mathcal{L}_{W_0}^{\rm HW}\Phi_{m0}=im\Phi_{m0}^{\rm HW}\;,
\end{align}
and their $\slr$ descendants obey
\begin{align}
    \label{eq:DescendantsHW}
	\Phi_{mn}^{\rm HW}=\mathcal{L}_{\hat{L}_-}^n\Phi_{m0}^{\rm HW}\;,\qquad
	\mathcal{L}_{\hat{L}_0}\Phi_{mn}^{\rm HW}=\pa{\hat{h}+n}\Phi_{mn}^{\rm HW}\;.
\end{align}
These $\slr$ towers of modes reproduce the exponentially decaying QNM overtones up to normalization:
\begin{align}
	\Phi_{mn}^{\rm HW}=2^{-\hat{h}+im}\br{\prod_{k=0}^{n-1}\hat{h}+k+im}\Phi_{mn}\;.
\end{align}
The exact value of the relative amplitude of successive QNM overtones is 
\begin{align}
    \label{eq:OvertoneHW}
	u_m=\frac{\Phi_{m1}^{\rm HW}(T,x,\phi)}{\Phi_{m0}^{\rm HW}(T,x,\phi)}
	=2e^{-T}\pa{\hat{h}\cosh{x}+im\sinh{x}}\;.
\end{align}
In the eikonal limit $\ab{m}\gg1$, we can approximate $\hat{h}\approx im\hat{\Omega}=\frac{im}{\hat{\lambda}}$, with $\hat{\lambda}=\hat{\lambda}_+$, simplifying this ratio to
\begin{align}
    \label{eq:EikonalRatioQNM}
	u_m\approx2im\hat{\Omega}e^{-T}\pa{\cosh{x}+\hat{\lambda}\sinh{x}}
	\sim e^{-T}\pa{\cosh{x}+\hat{\lambda}\sinh{x}}\;.
 \end{align}

In the next sections, we will want to compare the exact modes \eqref{eq:HWQNM} to their near-ring and geometric optics approximations near the (outer) bound orbit $x\approx\tilde{x}$.
Letting $\dt x=x-\tilde{x}$ and using $\hat{h}\stackrel{\ab{m}\gg1}{\approx}im\hat{\Omega}$, we find that  
\begin{align}
    e^{imx}\pa{\sinh{x}}^{\hat{h}}=e^{imx}e^{\hat{h}\log{\sinh{x}}}
    \stackrel{\ab{m}\gg1}{\approx}e^{imx+im\hat{\Omega}\log{\sinh{x}}}
    \stackrel{\ab{\dt x}\ll1}{\approx}c_0e^{-\frac{i}{2}\frac{m\hat{\Omega}}{\Lambda^2-1}\dt x^2},
\end{align}
so that very near the bound orbit, the exact QNM wave functions \eqref{eq:HWQNM} with large $m$ behave (up to a constant) as 
\begin{align}
    \label{eq:CompareExactNearRing}
	\Phi_{m0}^{\rm HW}(T,\dt x,\phi)\sim e^{-\frac{1}{2}T-im\hat{\Omega}\pa{T+\frac{1}{2}k\dt x^2}+im\phi}\;,\qquad
	k\equiv\frac{1}{\Lambda^2-1}\;.
\end{align}
A similar analysis shows that the anti-QNM overtones are $\slr$-descendants of the lowest-weight $n=0$ mode.

\subsection{Quasinormal modes in the near-ring region}
\label{subsec:NearRing}

In this section, we study the QNMs in the near-ring region of the warped AdS$_3$ metric \eqref{eq:WAdS3} and show that they form a shadow pair of highest-weight representations of an emergent near-ring $\slr_{\rm QN}$ symmetry, which differs from the geometrically realized conformal symmetry generated by the vector fields $\hat{L}_m$ given in \eqref{eq:Isometries}.

The near-ring region for waves on warped AdS$_3$ is defined in analogy with \eqref{eq:NearRingRegion} as
\begin{align}
    \label{eq:NearRingWaves}
    \text{NEAR-RING REGION:}\qquad
    \begin{cases}
        \ab{\dt r}\ll T_H
        \qquad\qquad\quad\ab{\dt x}\ll1
        &\qquad\text{(near-peak)}\;,\\
        \displaystyle\ab{\frac{m}{\omega_R}-\tilde{\lambda}}\ll\frac{1}{T_H}
        \qquad\,\ab{\frac{m}{\hat{\omega}_R}-\hat{\lambda}}\ll1
        &\qquad\text{(near-critical)}\;,\vspace{3pt}\\
        \displaystyle\frac{1}{\omega_R}\ll\frac{1}{T_H}
        \qquad\qquad\quad1\ll\hat{\omega}_R
        &\qquad\text{(high-frequency)}\;.
    \end{cases}
\end{align}
This defines a region of the phase space of waves on warped AdS$_3$, rather than simply a region of spacetime.
Still focusing on the outer photon shell, the wave potential \eqref{eq:RadialODE} in this near-ring region takes the simple form
\begin{align}
    V(\tilde{x}+\dt x)\approx\frac{\hat{\omega}_R^2}{\pa{\Lambda^2-1}^2}\dt x^2-\frac{2i\hat{\omega}_R\hat{\omega}_I}{\Lambda^2-1}\;,
\end{align}
and the radial ODE for $\psi(x)$ reduces to
\begin{align}
    \frac{1}{2\hat{\omega}_R}\br{\pd_{\dt x}^2+\frac{\hat{\omega}_R^2}{\pa{\Lambda^2-1}^2}\dt x^2}\psi(\dt x)=\frac{i\hat{\omega}_I}{\Lambda^2-1}\psi(\dt x)\;.
\end{align}
We may rewrite this in terms of $k=\pa{\Lambda^2-1}^{-1}>0$ as
\begin{align}
    \mathcal{H}\psi=i\hat{\omega}_I\psi\;,\qquad
    \mathcal{H}=\frac{1}{2k\hat{\omega}_R}\br{\pd_{\dt x}^2+k^2\hat{\omega}_R^2\dt x^2}\;.
\end{align}
This is precisely the time-independent Schr\"odinger equation for eigenstates $\psi$ of the upside-down harmonic oscillator with associated eigenvalues $i\hat{\omega}_I$, which are allowed to be imaginary because the boundary conditions are non-Hermitian.
Interestingly, in this example, the near-ring approximation for the imaginary part of the QNM spectrum is exact.

Following \cite{Hadar2022}, we now define the operators (recall that $T=\gamma_Lt$)
\begin{gather}
    \label{eq:SL2RQN}
    a_\pm=\frac{e^{\pm T}}{\sqrt{2k\hat{\omega}_R}}\pa{\mp\pd_{\dt x}-ik\hat{\omega}_R\dt x}\;,\qquad
    L_0=-\frac{i}{4}\pa{a_+a_-+a_-a_+}
	=\frac{i}{2}\mathcal{H}\;,\qquad
	L_\pm=\pm\frac{a_\pm^2}{2}\;.
\end{gather}
The $a_\pm$ generate the Heisenberg algebra $\br{a_+,a_-}=iI$, while the $L_m$ obey the exact $\slr_{\rm QN}$ commutation relations:
\begin{align}
    \br{L_0,L_\pm}=\mp L_\pm\;,\qquad
    \br{L_+,L_-}=2L_0\;.
\end{align}
Although these operators are defined everywhere, they are only useful in the near-ring region where $L_0$ is proportional to the Hamiltonian $\mathcal{H}$. 
The eigenstates of $L_0$ satisfy $L_0\psi_h=h\psi_h$, so we may identify $\hat{\omega}_I=-2h$, or equivalently, $\omega_I=-2\gamma_Lh$.
The mode ansatz $\Phi(T,x,\phi)=e^{-i\hat{\omega}T+im\phi}\psi(x)$ reduces in the near-ring region \eqref{eq:NearRingWaves} to
\begin{align}
    \Phi(T,\dt x,\phi)\approx e^{-i\hat{\omega}_RT+im\phi}\Phi_h(T,\dt x)\;,\qquad
    \Phi_h(T,\dt x)=e^{\hat{\omega}_IT}\psi_h(\dt x)
    =e^{-2hT}\psi_h(\dt x)\;.
\end{align}
Though the operator $L_0$ does not admit a normalizable ground state, it can still produce a discrete spectrum provided that boundary conditions are chosen appropriately.
The QNM boundary condition for the fundamental mode is equivalent to the highest-weight condition $L_+\Phi_h=0$, which admits two solutions with $h=\frac{1}{4}$ and $h=\frac{3}{4}$:
\begin{align}
    \Phi_\frac{1}{4}(T,\dt x)&=e^{-\frac{1}{2}T}\psi_\frac{1}{4}(\dt x)\;,
    &&\psi_\frac{1}{4}(\dt x)=e^{-\frac{i}{2}k\hat{\omega}_R\dt x^2}\;,\\
    \Phi_\frac{3}{4}(T,\dt x)&=e^{-\frac{3}{2}T}\psi_\frac{3}{4}(\dt x)\;,
    &&\psi_\frac{3}{4}(\dt x)=\dt x\,e^{-\frac{i}{2}k\hat{\omega}_R\dt x^2}\;.
\end{align}
The action of $\slr_{\rm QN}$ then produces a tower of descendants that are precisely the higher overtones:
\begin{align}
    \Phi_{h,N}(T,\dt x)=L_-^N\Phi_h(T,\dt x)
    =e^{-2(h+N)T}\psi_{h+N}(\dt x)
    \propto e^{-2(h+N)T}D_{2(h+N)-\frac{1}{2}}\pa{\sqrt{2ik\hat{\omega}_R}\dt x}\;,
\end{align}
where $D_n(t)$ denotes the $n^\text{th}$ parabolic cylinder function.
Since at the edges $\dt x\to\pm\infty$ of the near-peak region,
\begin{align}
    D_n\pa{\sqrt{2ik\hat{\omega}_R}\dt x}\stackrel{\dt x\to\pm\infty}{\approx}\dt x^ne^{-\frac{i}{2}k\hat{\omega}_R\dt x^2}\;,
\end{align}
and recalling that $\hat{\omega}_R=m\hat{\Omega}$, it follows that the $n^\text{th}$ overtone behaves near the edges as
\begin{align}
    \label{eq:WaveFunctions}
    \Phi_{mn}(T,\dt x,\phi)\stackrel{\dt x\to\pm\infty}{\sim}e^{-\pa{n+\frac{1}{2}}T}\dt x^ne^{-i\hat{\omega}_R\pa{T+\frac{1}{2}k\dt x^2}+im\phi}\;,
\end{align}
which agrees with the highest-weight QNMs \eqref{eq:CompareExactNearRing} in the near-ring region \eqref{eq:NearRingWaves}.
Likewise, imposing a lowest-weight condition would reproduce the $n=0$ anti-QNM \eqref{eq:CompareExactAQNMNearRing}, whose $L_+$ descendants are the near-ring anti-QNM overtones
\begin{align}
	\tilde{\Phi}_{mn}(T,\dt x,\phi)\stackrel{\dt x\to\pm\infty}{\sim}e^{+\pa{n+\frac{1}{2}}T}\dt x^ne^{-i\hat{\omega}_R\pa{T-\frac{1}{2}k\dt x^2}+im\phi}\;.
\end{align}

Now we would like to compare the emergent $\slr_{\rm QN}$ symmetry of the near-ring region with the exact spacetime isometry $\slr_{\rm ISO}$.
We will only compare their actions on the QNMs within the near-ring region in which the $\slr_{\rm QN}$ symmetry actually emerges.
In that region, $\slr_{\rm QN}$ acts on functions via the differential operators $L_\pm$ defined in \eqref{eq:SL2RQN}, while $\slr_{\rm ISO}$ acts on them via the Lie derivative taken along its vector field generators \eqref{eq:Isometries}.

From \eqref{eq:DescendantsHW}, we see that $\slr_{\rm ISO}$ acts on the QNM overtone number as $n\to n+1$, while from the preceding discussion we see that $\slr_{\rm QN}$ acts on it as $n\to n+2$.
It thus follows that within the near-ring region,
\begin{align}
    \label{eq:Squaring}
    L_\pm\Phi_{mn}\propto\mathcal{L}_{\hat{L}_\pm}^2\Phi_{mn}\;.
\end{align}
Similarly, $L_0\Phi_{mn}=\frac{i}{2}\mathcal{H}\Phi_{mn}=-\frac{1}{2}\hat{\omega}_I\Phi_{mn}$, while using $\Phi_{mn}\sim e^{-i\hat{\omega}T+im\phi}$, we have $\mathcal{L}_{\hat{L}_0}\Phi_{mn}=-\pd_T\Phi_{mn}=i\hat{\omega}\Phi_{mn}$ and $\mathcal{L}_{W_0}\Phi_{mn}=\pd_\phi\Phi_{mn}=im\Phi_{mn}$, so that $\pa{\mathcal{L}_{\hat{L}_0}-\hat{\Omega}\mathcal{L}_{W_0}}\Phi_{mn}=i\pa{\hat{\omega}_R+i\hat{\omega}_I-m\hat{\Omega}}\Phi_{mn}=-\hat{\omega}_I\Phi_{mn}$. Hence,
\begin{align}
    L_0\Phi_{mn}=\frac{1}{2}\pa{\mathcal{L}_{\hat{L}_0}-\hat{\Omega}\mathcal{L}_{W_0}}\Phi_{mn}.
\end{align}

\subsection{Quasinormal modes from geometric optics}
\label{subsec:GeometricOptics}

In this section, we reproduce some results from sections~\ref{subsec:Spectrum} and \ref{subsec:ConformalSymmetry} using the geometric optics approximation.
Waves with large frequencies compared to the local curvature scale take the approximate form 
\begin{align}
    \Phi\approx Ae^{iS}\;,
\end{align}
with a rapidly varying phase $S(x^\mu)$ and a slowly varying amplitude $A(x^\mu)$.
In terms of the gradient
\begin{align}
    p_\mu=\pd_\mu S\;,
\end{align} 
which is normal to the wavefronts of constant $S(x^\mu)$, the wave equation takes the form
\begin{align}
    \label{eq:ShortWave}
    -p^\mu p_\mu A+i\pa{2p^\mu\nabla_\mu A+\nabla_\mu p^\mu A}+\nabla^2A=0\;.
\end{align}
In the geometric optics approximation, this equation is solved order-by-order in inverse powers of $p$.
At leading order,
\begin{align}
    p_\mu p^\mu=0\;,\qquad
    p^\mu\nabla_\mu p_\nu=0\;,
\end{align}
so that $p$ is required to be tangent to a null geodesic.
At subleading order, the parallel transport of the amplitude along the congruence is related to the expansion $\theta=\nabla_\mu p^\mu$ as 
\begin{align}
    \label{eq:GeometricAmplitude}
    p^\mu\pd_\mu\log{A}=-\frac{1}{2}\theta\;.
\end{align}
This implies that the expansion controls the exponential behavior of the amplitude with respect to affine parameter:
\begin{align}
    \pd_s\log{A}=-\frac{1}{2}\theta
    \qquad\implies\qquad
    A\sim A_0e^{-\frac{1}{2}\theta s}\;.
\end{align}
According to \eqref{eq:GeometricAmplitude}, any quantity that does not vary along the null congruence can be used to build towers of approximate solutions to the wave equation given some seed solution.
More precisely, if a function $u(x^\mu)$ satisfies 
\begin{align}
    \label{eq:ConservedQuantity}
    p^\mu\pd_\mu u=0
\end{align}
along the congruence, and moreover if
\begin{align}
    \Phi_0\approx A_0e^{iS}
\end{align}
is an approximate (to leading and subleading order) solution to \eqref{eq:ShortWave}, then it follows from \eqref{eq:GeometricAmplitude} that
\begin{align}
    \Phi_n\approx u^nA_0e^{iS}
\end{align}
is also an approximate solution to the wave equation.
In other words, if $A_0$ is an amplitude solving \eqref{eq:GeometricAmplitude} along a congruence with tangent $p$, then so is $A_n=u^nA_0$.
As we will show, the minimal solution to \eqref{eq:GeometricAmplitude} for homoclinic rays in the near-ring region \eqref{eq:NearRingWaves} takes the form $A_0=e^{-\frac{1}{2}T}\sqrt{\sinh x}$, and according to \eqref{eq:GeodesicDeviation}--\eqref{eq:LyapunovExponent}, the quantity
\begin{align}
    \label{eq:Quantity}
    u=e^{-T}\dt x
\end{align}
is constant on the unstable homoclinic orbit and obeys \eqref{eq:ConservedQuantity} in the near-ring region.
Geometric optics therefore produces a family of solutions associated to the same bound photon orbit but differing in the amplitude:
\begin{align}
    \label{eq:WKBgeneral}
    \Phi_{mn}(T,\dt x,\phi)\approx e^{-\pa{n+\frac{1}{2}}T}\pa{\dt x}^ne^{iS(T,\dt x,\phi)}\;.
\end{align}
As we will now show, this expression precisely reproduces the near-ring expression \eqref{eq:WaveFunctions} for the QNMs.

Null geodesics in \eqref{eq:SimpleWAdS3} have a conserved energy $\hat{\omega}=-p_T$ and angular momentum $m=p_\phi$.
Their momentum is
\begin{align}
	p_\mu=\pd_\mu S\;,\qquad
	S=-\hat{\omega}T+m\phi\pm_x\int^x\sqrt{V(x')}\ed x'\;,
\end{align}
where $V(x)$ is the radial potential \eqref{eq:RadialODE}, $\pm_x=\sign{p_x}$, and $S$ is a Hamilton-Jacobi principal function.
Here, we focus on unstable homoclinic rays with impact parameter $m/\hat{\omega}=\hat{\lambda}=\hat{\Omega}^{-1}$ given in \eqref{eq:Criticality}.
Their momentum and phase are
\begin{align}
    \label{eq:CriticalS}
    \hat{p}_\mu=\pd_\mu \hat{S}\;,\qquad
	\hat{S}&=\hat{\omega}\br{-T+\hat{\lambda}\phi+\int_{\tilde{x}}^x\pa{\frac{1}{\tanh{x'}}+\hat{\lambda}}\ed x'}
	=\hat{\omega}\br{-T+\hat{\lambda}\phi+\hat{\lambda}x+\log\pa{\sinh{x}}+C}\;,
\end{align}
where the integration constant $C$ is an irrelevant phase shift.
Because these geodesics asymptote to the peak of their radial potential $V(x)$ (located at a double root $x=\tilde{x}$, which they cannot cross), they form congruences with precisely the right behavior to approximate wave solutions obeying QNM boundary conditions (namely, purely ingoing at the horizon and purely outgoing at the boundary).
The expansion of these congruences is
\begin{align}
	\theta=\nabla^2\hat{S}
	=-\frac{\hat{\omega}}{\ell^2}\;.
\end{align}
The subleading equation \eqref{eq:GeometricAmplitude} for the wave amplitude $A(T,x)$ takes the explicit form
\begin{align}
	\br{\pa{\frac{1}{\tanh{x}}+\hat{\lambda}}\pd_x+\pa{1+\frac{\hat{\lambda}}{\tanh{x}}}\pd_T-\frac{1}{2\sinh^2{x}}}A(T,x)=0\;.
\end{align}
It admits a minimal solution
\begin{align}
    A_0(T,x)=e^{-\frac{1}{2}T}\sqrt{\sinh{x}}\;,
\end{align}
which generates an infinite family of solutions of the form
\begin{align}
    \label{eq:OvertoneRatio}
	A_n(T,x)=u^n(T,x)A_0(T,x)\;,\qquad
	u(T,x)=e^{- T}\pa{\cosh{x}+\hat{\lambda}\sinh{x}}\;.
\end{align}
Recalling that $\hat{\omega}=m\hat{\Omega}$ for homoclinic rays, the eikonal approximation to the QNMs is
\begin{align}
	\label{eq:ApproximateQNM}
	\Phi_{mn}(T,x,\phi)\approx A_n(T,x)e^{i\hat{S}(T,x,\phi)}
	=e^{-nT}\pa{\cosh{x}+\hat{\lambda}\sinh{x}}^ne^{-\pa{\frac{1}{2}+im\hat{\Omega}}T+im\phi+ imx}\pa{\sinh{x}}^{\frac{1}{2}+im\hat{\Omega}}\;.
\end{align}
In particular, setting $n=0$ in this formula results in
\begin{align}
	\Phi_{m0}(T,x,\phi)\approx A_0(T,x)e^{i\hat{S}(T,x,\phi)}
	=e^{-\pa{\frac{1}{2}+im\hat{\Omega}}T+im\phi+ imx}\pa{\sinh{x}}^{\frac{1}{2}+im\hat{\Omega}}\;,
\end{align}
which precisely reproduces the eikonal limit \eqref{eq:EikonalQNM}  of the fundamental QNM.
Likewise, the geodesic approximation \eqref{eq:ApproximateQNM} predicts that successive overtones differ by the ratio $u(T,x)$ given in \eqref{eq:OvertoneRatio}, which also matches the eikonal limit of the exact ratio given in \eqref{eq:EikonalRatioQNM} for QNMs.
We note that to first order in the near-ring region,
\begin{align}
    u(T,\tilde{x}+\dt x)\approx-\frac{e^{- T}\dt x}{\sqrt{\Lambda^2-1}}\;,
\end{align}
so that in this region of phase space, these quantities take the simple form $u\sim e^{-T}\dt x$, in accord with \eqref{eq:Quantity}.

When the spacetime admits an $\slr$ isometry (as is the case here), the geometric optics construction of the overtone wavefunctions has a clear explanation in terms of this symmetry.
On a homoclinic trajectory with critical momentum $\hat{p}$, the Hamiltonian functions \eqref{eq:Functions} take the approximate form in the near-ring region $x=\tilde{x}+\dt x$
\begin{align}
    \hat{L}_+(\hat{p})=0\;,\qquad
    \hat{L}_-(\hat{p})=2\hat{\omega}e^{- T}\pa{\cosh{x}+\hat{\lambda}\sinh{x}}
    \propto u(T,x)\;.
\end{align}
This is equivalent to the statement that the phase function $\Psi_0=e^{i\hat{S}}$ is highest-weight with respect to $\slr_{\rm ISO}$, and we could equivalently express \eqref{eq:OvertoneRatio} as 
\begin{align}
    \label{eq:Lowering}
    A_n(T,x)\propto\hat{L}_-^n(T,x,\hat{p})A_0(T)\;.  
\end{align}
This equation states that to generate higher overtone amplitudes based on a homoclinic congruence, one simply multiplies the seed amplitude by the Hamiltonian function $\hat{L}_-(x,\hat{p})$.
This is of course the geometric optics analogue of the exact statement that the higher-overtone wavefunctions are conformal descendants,
\begin{align}
    \Phi_{mn}\propto\mathcal{L}_{\hat{L}_-}^n\Phi_{m0}\;. 
\end{align}
This follows because, to leading order in the geometric optics approximation, 
\begin{align}
    \hat{L}_\mp^\mu\pd_\mu\pa{A_0e^{i\hat{S}}}=iA_0\pa{\hat{L}_\mp^\mu\hat{p}_\mu}e^{i\hat{S}}+\text{subleading}\;.
\end{align}
The constructions \eqref{eq:ApproximateQNM} and \eqref{eq:Lowering} are equivalent because they both rely on solutions to the equation $p^\mu\pd_\mu L(x,p)=0$.
The quantity $\hat{L}_-(x,p)=\hat{L}^\mu_- p_\mu$ solves this equation for any geodesic congruence by virtue of the Killing equation, while $u(T,x)$ satisfies it \textit{specifically for the homoclinic trajectory} $\hat{p}$ to which the QNMs are associated.

\section{Observable conformal symmetry of the ``photon ring''}
\label{sec:Symmetries}

In this section, we define the observable conformal symmetry $\slr_{\rm PR}$ of the ``photon ring'' and explore its relationship with the spacetime isometry group $\slr_{\rm ISO}$ of the warped AdS$_3$ metric \eqref{eq:WAdS3}. 

\subsection{Successive images are related by a dilation}

Black hole images in three dimensions are somewhat trivial, since the observer's screen is just a line, rather than a plane, and the image of the (outer) bound orbit is a single critical point, rather than a (closed) critical curve.
However, this simple case is still rich enough to illustrate the $\slr_{\rm PR}$ construction recently applied to Schwarzschild and Kerr black holes \cite{Hadar2022}.
Since the WAdS$_3$ metric \eqref{eq:WAdS3} is not of direct astrophysical interest, we treat it telegraphically and refer the reader to \cite{Hadar2022} for a more detailed analysis.

The phase space $\Gamma$ of null geodesics in the geometry \eqref{eq:SimpleWAdS3} is four-dimensional, with coordinates $(x,\phi,p_x,p_\phi)$, canonical symplectic form $\Omega=\ed p_x\wedge\ed x+\ed p_\phi\wedge\ed\phi$, and Hamiltonian (obtained by solving $g^{\mu\nu}p_\mu p_\nu=0$ for $H=-p_t$)
\begin{align}
	H=-\frac{p_\phi}{\tanh{x}}-\sqrt{p_x^2+\frac{p_\phi^2}{\Lambda^2\sinh^2{x}}}\;.
\end{align}
The (outer) bound orbit has $(x,p_x)=(\tilde{x},0)$, and critical energy
\begin{align}
	\tilde{H}=\frac{p_\phi}{\hat{\lambda}}
	=-p_\phi\sqrt{1-\frac{1}{\Lambda^2}}
	>0\;,
\end{align}
which is positive since $p_\phi<0$ because the orbit is counter-rotating.
This dynamical system is integrable and therefore admits a canonical transformation to action-angle variables $(x,\phi,p_x,p_\phi)\to(T,\Phi,H,L)$ given by
\begin{align}
	\label{eq:ActionAngle}
	\ed T=\frac{H+L\coth{x}}{\sqrt{\mathcal{V}(x)}}\ed x\;,\qquad
	\ed\Phi=\ed\phi+\br{\coth{x}\pa{H+L\coth{x}}-\frac{L}{\Lambda^2\sinh^2{x}}}\frac{\ed x}{\sqrt{\mathcal{V}(x)}}\;,\qquad
	L=p_\phi\;,
\end{align}
where the radial geodesic potential is
\begin{align}
    \mathcal{V}(x)=\pa{H+L\coth{x}}^2-\frac{L^2}{\Lambda^2\sinh^2{x}}\;.
\end{align}
As expected, this canonical transformation preserves the symplectic form: $\Omega=\ed H\wedge\ed T+\ed L\wedge\ed\Phi$.

The action-angle variables trivialize the dynamics:
\begin{align}
	\dot{H}=\cu{H,H}=0\;,\qquad
	\dot{L}=\cu{L,H}=0\;,\qquad
	\dot{\Phi}=\cu{\Phi,H}=0\;,\qquad
	\dot{T}=\cu{T,H}=1\;.
\end{align}
The first two equations indicate that the phase space $\Gamma$ foliates into superselection sectors of fixed $(H,L)$, which are conserved momenta.
The third equation implies that a photon with initial coordinates $(x_s,\phi_s,H,L)$ evolves to final coordinates $(x_o,\phi_o,H,L)$ such that
\begin{align}
    \label{eq:AzimuthSwept}
	\Delta\phi=\phi_o-\phi_s
	=\fint_{\phi_s}^{\phi_o}\ed\phi
	=-\fint_{x_s}^{x_o}\br{\coth{x}\pa{H+L\coth{x}}-\frac{L}{\Lambda^2\sinh^2{x}}}\frac{\ed x}{\sqrt{\mathcal{V}(x)}}\;,
\end{align}
where the slash indicates that the integral is to be evaluated along the photon trajectory.
The last equation identifies $T$ as the variable conjugate to energy, i.e., time.
Hence, the time elapsed along a geodesic as it evolves from $(x_s,\phi_s,H,L)$ to $(x_o,\phi_o,H,L)$ is
\begin{align}
    \label{eq:TimeLapse}
	T=\fint_{x_s}^{x_o}\frac{H+L\coth{x}}{\sqrt{\mathcal{V}(x)}}\ed x\;.
\end{align}
Equations \eqref{eq:AzimuthSwept} and \eqref{eq:TimeLapse} represent the solution to the null geodesic equation in the warped AdS$_3$ metric \eqref{eq:SimpleWAdS3}, reduced to a problem of quadratures.
These integrals can be evaluated in terms of elliptic integrals \cite{Kapec2020}.

Since we are interested in optical images of sources far from the black hole, here we focus on geodesics that begin and end at the boundary $x=0$ and therefore always remain outside the bound orbit at $x=\tilde{x}$.
A distant observer at large radius $x_o\to0$ near the boundary receives these geodesics with impact parameter
\begin{align}
	\lambda=\frac{L}{H}
	>\hat{\lambda}
	=-\frac{1}{\sqrt{1-\frac{1}{\Lambda^2}}}\;.
\end{align}
The radius of closest approach is reached when the radial momentum $p_x=\pm\sqrt{\mathcal{V}(x)}$ vanishes.
This occurs at the smallest root of the radial potential $\mathcal{V}(x)$, namely
\begin{align}
	x_{\rm min}=\arctanh\br{\frac{\lambda\sqrt{\Lambda^2+\lambda^2\pa{1-\Lambda^2}}-\lambda\Lambda^2}{\Lambda^2+\lambda^2}}
	\le\tilde{x}
	=\arctanh\sqrt{1-\frac{1}{\Lambda^2}}\;.
\end{align}
Geodesics with $\hat{H}=0$ are homoclinic and asymptote to the closed photon orbit at $x=\tilde{x}$ in the far past and/or future.
Their impact parameter $\lambda=\hat{\lambda}$ defines the ``critical point'' on the image line.
Points to the right of this critical point can be illuminated, while the observer screen to the left is dark.

It is possible to define on the phase space $\Gamma$ an action of a conformal group $\slr_{\rm PR}$ with generators
\begin{align}
    \label{eq:SL2RPR}
	H_+=\hat{H}\;,\qquad
	H_0=-TH_+\;,\qquad
	H_-=T^2H_+\;,
\end{align}
which obey the correct algebra since $(T,H)$ are canonically conjugate.
These transformations correspond to the finite $\slr\cong\mathsf{Sp}(2,\mathbb{R})$ symplectomorphisms (App.~\ref{app:Flows})
\begin{align}
	\pa{T,\hat{H}}\longrightarrow\pa{\frac{aT+b}{cT+d},\pa{cT+d}^2\hat{H}}\;,\qquad
	ad-bc=1\;,
\end{align}
which leave the photon shell invariant---it is, in fact, the unique $\slr_{\rm PR}$-invariant submanifold of phase space.

$\slr_{\rm PR}$ also has an interesting action on the image line. 
Since it commutes with the $\mathsf{U}(1)$ generated by $L$, it acts within superselection sectors $\Gamma_L$ of fixed angular momentum.
However, it does modify the energy (or photon color) $H=H_++\tilde{H}$.
It therefore also modifies the impact parameter and the radius of closest approach for each geodesic.
Indeed, the ``photon ring'' is an attractive fixed point for the phase space flow generated by $e^{-\alpha H_0}$, under which
\begin{align}
	\label{eq:FiniteDilation}
	\hat{H}(0)\to\hat{H}(\alpha)=e^{-\alpha}\hat{H}(0)\;.
\end{align}
For large $\alpha$, $\hat{H}$ becomes small while $T \to \infty$.

For a homoclinic orbit with $H_+=0$, expanding the first relation in \eqref{eq:ActionAngle} for small $\dt x=x-\tilde{x}<0$ yields
\begin{align}
	\ed T\approx\ed\log\dt x
	\qquad\Longrightarrow\qquad
	\dt x\approx\dt x_0e^T\;.
\end{align}
Since the point of closest approach to the photon shell is given by
\begin{align}
    \dt x_{\rm min}^2=\frac{2\Lambda^2}{\hat{\lambda}}\frac{\hat{H}}{L}+\ldots
\end{align}
to leading order as $\hat{H}\to0$, $\log{\dt x_{\rm min}}$ decreases linearly under the $\slr_{\rm PR}$ dilations \eqref{eq:FiniteDilation} and
\begin{align}
    \pd_\alpha\log\dt x_{\rm min}=-\frac{1}{2}\;.
\end{align}
Expanding \eqref{eq:AzimuthSwept} for small $\dt x_{\rm min}$ shows that near the photon shell,
\begin{align}
    \Delta\phi=\frac{1}{\hat{\lambda}}\log{\dt x_{\rm min}}\;,
\end{align}
so that the number of orbits around the black hole $w=\Delta\phi/2\pi$ diverges linearly under dilations:
\begin{align}
    \pd_\alpha w=-\frac{1}{4\pi\hat{\lambda}}
    =\frac{1}{2\gamma}\,.
\end{align}
We would now like to understand the effect of the dilations \eqref{eq:FiniteDilation} on images of a source in the geometry \eqref{eq:SimpleWAdS3}.
We assume there is a source star located at $(x_s,\phi_s)$ emitting light of all colors isotropically, as well as a telescope at $(x_o,\phi_o)$.
There are an infinite number of (colored) geodesics connecting the source to the observer, each distinguished by the winding number around the black hole.
Since these geodesics share the same endpoints, they also share the same net angular shift $\Delta\phi$ modulo $2\pi$.
Geodesics with $\ab{w}>\frac{1}{2}$ contribute to the observed photon ring.

Consider null geodesics of fixed $p_\phi=L$.
Demanding that the geodesics connect a source point $(x_s,\phi_s)$ with an observer point $(x_o,\phi_o)$ cuts $\Gamma_L$ down to a discrete set of geodesics $\Gamma_{\rm obs}$ labeled by the (integer part of their) winding number $\lfloor w\rfloor$.
In the limit $w\to\infty$, the finite dilation
\begin{align}
   D_0=e^{-2\gamma H_0}
\end{align}
relates geodesics with successive winding numbers $w\to w+1$.
The products $D_0^k$ for $k\in\mathbb{Z}$ constitute a ($w$-independent) discrete subgroup of $\slr_{\rm PR}$ mapping $\Gamma_{\rm obs}$ onto itself.
This emergent discrete scaling symmetry generates successive subring images in the ``photon ring'' that asymptotically approach the critical point.

\subsection{\texorpdfstring{$\slr_{\rm PR}$ and $\slr_{\rm ISO}$}{Photon ring SL(2,R) = Isometry SL(2,R)}}

When the black hole reaches zero temperature, $T_H=0$, the metric \eqref{eq:WAdS3} reduces to
\begin{align}
	\label{eq:ZeroTemperatureMetric}
	ds^2=\ell^2\br{-r^2\ed t^2+\frac{\ed r^2}{r^2}+\Lambda^2\pa{\ed\phi+r\ed t}^2}\;.
\end{align}
This spacetime admits an isometry group $\slr_{\rm ISO}^0$ generated by Killing vectors
\begin{align}
    \label{eq:SL2R0}
    \hat{H}_+=\pd_t\;,\qquad
    \hat{H}_0=t\pd_r-r\pd_r\;,\qquad
    \hat{H}_-=\pa{t^2+\frac{1}{r^2}}\pd_t-2tr\pd_r-\frac{2}{r}\pd_\phi\;,
\end{align}
which act on the six-dimensional phase space $(t,r,\phi,p_t,p_r,p_\phi)$ via $\slr_{\rm ISO}^0$-generating functions $\hat{H}_n=-\hat{H}_n^\mu p_\mu$.
Like their finite-temperature analogues $\hat{L}_n$, these functions all commute with the constrained Hamiltonian
\begin{align}
    \label{eq:ZeroHamiltonianEPS}
    \mathcal{H}=\frac{1}{2}g^{\mu\nu}p_\mu p_\nu
    =\frac{1}{2\ell^2}\br{r^2p_r^2-\frac{\pa{p_t-rp_\phi}^2}{r^2}+\frac{p_\phi^2}{\Lambda^2}}
    \equiv0\;,
\end{align}
so they are conserved charges along geodesics, and they have the same Casimir
\begin{align}
    \mathcal{C}=-\hat{H}_0^2+\frac{\hat{H}_+\hat{H}_-+\hat{H}_-\hat{H}_+}{2}
    =-\hat{L}_0^2+\frac{\hat{L}_+\hat{H}_-+\hat{L}_-\hat{L}_+}{2}
    =-\pa{1-\frac{1}{\Lambda^2}}p_\phi^2\;.
\end{align}

In the four-dimensional phase space $(r,\phi,p_r,p_\phi)$, null geodesic motion is generated by the Hamiltonian $H=-p_t$, which is obtained by solving the dynamical constraint $\mathcal{H}=0$ to find
\begin{align}
    \label{eq:ZeroHamiltonianRPS}
    H=-r\pa{p_\phi+\sqrt{r^2p_r^2+\frac{p_\phi^2}{\Lambda^2}}}\;.
\end{align}
As reviewed in App.~\ref{app:Reduction}, the coordinate time $T(r,\phi,p_r,p_\phi)$ elapsed along geodesic flow is the function that obeys $\cu{T,H}=1$, which for this Hamiltonian gives (up to an integration constant $\mathcal{T}_0$)
\begin{align}
    \label{eq:ZeroTime}
    T=-\frac{rp_r}{H}-\mathcal{T}_0\;.
\end{align}
Under phase space reduction at time $t=t_0$, the functions $\hat{H}_n=-\hat{H}_n^\mu p_\mu$ defined from \eqref{eq:SL2R0} reduce to
\begin{align}
    \label{eq:SL2RPRandISO}
    \hat{h}_+=H\;,\qquad
    \hat{h}_0=t_0H+rp_r\;,\qquad
    \hat{h}_-=\pa{t_0^2+\frac{1}{r^2}}H+2t_0rp_r+\frac{2}{r}p_\phi\;,
\end{align}
where $t_0$ is now just a constant.
We note that the functions $\hat{h}_\pm$ are \textit{no longer} conserved in four-dimensional phase space, since they do not commute with the Hamiltonian $H$ and therefore change the energy (or photon color) of a null geodesic.
We now fix $\mathcal{T}_0=t_0$ and set
\begin{align}
    T=-\frac{rp_r}{H}-t_0\;,\qquad
    \mathcal{C}\equiv-\hat{h}_0^2+\frac{\hat{h}_+\hat{h}_-+\hat{h}_-\hat{h}_+}{2}
    =-\pa{1-\frac{1}{\Lambda^2}}p_\phi^2\;.
\end{align}
Inverting these relations allows us to rewrite \eqref{eq:SL2RPRandISO} as
\begin{align}
    \hat{h}_+=H\;,\qquad
    \hat{h}_0=-TH\;,\qquad
    \hat{h}_-=T^2H+\frac{\mathcal{C}}{H}\;,
\end{align}
which shows that the phase space reduction of $\slr_{\rm ISO}^0$ reproduces the general form of $\slr_{\rm PR}$ given in \eqref{eq:CasimirChoice}.

As discussed in App.~\ref{app:Reduction}, 
the $\slr_{\rm PR}$ construction actually leads to a two-parameter family of phase space $\slr$'s labeled by a choice of critical energy $\tilde{H}$ and Casimir $\mathcal{C}$ in the definition of $H_-$.
The choice of $\tilde{H}$ determines the energy hypersurface that $\slr_{\rm PR}$ drives geodesics onto, whereas the ambiguity in the choice of $\mathcal{C}$ included in the generator $H_-$ does not affect the scaling into the photon shell, which only depends on the dilations $H_0$.
Here, we see that this Casimir takes the particular form $\mathcal{C}=-\pa{1-\Lambda^{-2}}p_\phi^2$ for $\slr_{\rm ISO}^0$.
In other words, the phase space reduction of $\slr_{\rm ISO}^0$ recovers the $\slr_{\rm PR}$ in \eqref{eq:SL2RPR} up to a shift $H\to\hat{H}=H-\tilde{H}$ and the inclusion of a specific Casimir.
We conclude that $\slr_{\rm PR}$ is a simple deformation of $\slr_{\rm ISO}^0$ within the more general family defined by \eqref{eq:CasimirChoice}.

It is well-known that the zero-temperature metric \eqref{eq:ZeroTemperatureMetric} can be mapped into the finite-temperature metric \eqref{eq:WAdS3} via the coordinate transformation
\begin{align}
    t\to\frac{e^{2\pi T_Ht}\pa{r+2\pi T_H}}{\sqrt{\pa{r+2\pi T_H}^2-\pa{2\pi T_H}^2}}\;,\qquad
    r\to\frac{\sqrt{\pa{r+2\pi T_H}^2-\pa{2\pi T_H}^2}}{2\pi T_He^{2\pi T_Ht}}\;,\qquad
    \phi\to\phi+\frac{1}{2}\log\pa{\frac{r}{r+4\pi T_H}}\;.
\end{align}
This transformation maps $\hat{H}_\pm\to-\hat{L}_\mp$ and $\hat{H}_0\to-\hat{L}_0$.
This demonstrates that the finite-temperature isometry group $\slr_{\rm ISO}$ and the observationally relevant $\slr_{\rm PR}$ given in \eqref{eq:SL2RPR} are both members of the two-parameter family of phase space $\slr$'s defined by \eqref{eq:CasimirChoice}.
The observational $\slr_{\rm PR}$ can be thought of as a particularly useful deformation of $\slr_{\rm ISO}$ within this family, since it provides an organizing principle for black hole images.

\section{Quantum Ruelle resonances = classical Lyapunov exponents}
\label{sec:Matching}

We saw in section~\ref{sec:WarpedAdS3} that the short-wave spectrum of QNMs in the geometry \eqref{eq:WAdS3},
\begin{align}
    \label{eq:WAdS3bis}
	ds^2=\ell^2\br{-r\pa{r+4\pi T_H}\ed t^2+\frac{\ed r^2}{r\pa{r+4\pi T_H}}+\Lambda^2\br{\ed\phi+\pa{r+2\pi T_H}\ed t}^2}\;,
\end{align}
is controlled by its photon shell.
It is well-known that this metric has left and right temperatures \cite{Guica2009,Bredberg2010} 
\begin{align}
	\label{eq:Temperatures}
	T_L=\frac{1}{2\pi}\;,\qquad
	T_R=T_H\;.
\end{align}
An ingoing scalar mode $\Phi(t,r,\phi)=e^{-i\omega t+im\phi}\psi^{\rm in}(r)$ in the background \eqref{eq:WAdS3bis} has the leading boundary behavior \eqref{eq:BoundaryIngoingModes}
\begin{align}
	\psi^{\rm in}(r)&\stackrel{r\to\infty}{\approx}\frac{\Gamma(2\beta)\Gamma\br{1-i\pa{\hat{\omega}+m}}}{\Gamma\pa{\frac{1}{2}+\beta-im}\Gamma\pa{\frac{1}{2}+\beta-i\hat{\omega}}}\pa{4\pi T_H}^{\frac{1}{2}-\beta-\frac{i}{2}\pa{\hat{\omega}+m}}r^{-\frac{1}{2}+\beta}+\pa{\beta\to-\beta}\;.
\end{align}
Resonances occur when one of these two terms vanishes.
Recalling that $\hat{\omega}=\omega/(2\pi T_H)$, this happens when
\begin{align}
	\frac{1}{2}\pm\beta-\frac{i\omega}{2\pi T_H}=-n\;,
	\qquad\text{or}\qquad
	\frac{1}{2}\pm\beta-im=-n\;.
\end{align}
The two different signs correspond to the two different bound photon orbits: the $+$ sign is for the inner photon shell at $r=\tr_-$ and the $-$ sign is for the the outer photon shell at $r=\tr_+$.
The resonances are thus
\begin{align}
    \label{eq:SpectrumMatching}
	\text{Inner shell:}\quad
	\begin{cases}
		\omega=-2\pi T_Hi\pa{n+\frac{1}{2}+\beta},\\
		m=-i\pa{n+\frac{1}{2}+\beta},
	\end{cases}
	\qquad\qquad
	\text{Outer shell:}\quad
	\begin{cases}
		\omega=-2\pi T_Hi\pa{n+\frac{1}{2}-\beta},\\
		m=-i\pa{n+\frac{1}{2}-\beta}.
	\end{cases}
\end{align}
As pointed out by \cite{Bredberg2010,Chen:2010qm,Li:2010sv}, this spectrum of resonances resembles that of a CFT$_2$ with the same left and right temperatures \eqref{eq:Temperatures} and a peculiar identification of momenta.
In a finite-temperature CFT$_2$, a primary operator with conformal weights $h_L$ and $h_R$ has a thermal two-point function $\av{[O(t,x),O(t',x')]}_{\beta}$ with Fourier transform \cite{Birmingham2002,Birmingham2003}
\begin{align}
	\label{eq:FeynmanProp}
	G(p_L,p_R)\sim \Gamma\pa{h_L-\frac{ip_L}{2\pi T_L}}\Gamma\pa{h_L+\frac{ip_L}{2\pi T_L}}\Gamma\pa{h_R-\frac{ip_R}{2\pi T_R}}\Gamma\pa{h_R+\frac{ip_R}{2\pi T_R}}\;,
\end{align}
and the poles of the retarded correlator $G_R(p_L,p_R)$ correspond to the poles in the lower half-plane:
\begin{align}
	\label{eq:RuelleResonances}
	\begin{cases}
		p_L=-2\pi iT_L(n+h_L)\;,\\
		p_R=-2\pi iT_R(n+h_R)\;.
	\end{cases}
\end{align}
Consider a scalar field with conformal weights $h_L=h_R=\frac{1}{2}-\beta$.
If the CFT$_2$ has the temperatures \eqref{eq:Temperatures} arising from the warped metric \eqref{eq:WAdS3bis}, then its spectrum of Ruelle resonances \eqref{eq:RuelleResonances} becomes
\begin{align}
	\begin{cases}
		p_L=-i\pa{n+\frac{1}{2}-\beta}\;,\\
		p_R=-2\pi iT_H\pa{n+\frac{1}{2}-\beta}\;,
	\end{cases}
\end{align}
which exactly matches the QNM spectrum associated with the outer photon shell in \eqref{eq:SpectrumMatching}, provided that we identify
\begin{align}
	(p_L,p_R)=(m,\omega)\;.
\end{align}
This is precisely the identification previously made in Eq.~(48) of \cite{Chen:2010qm} and Eq.~(2.13) of \cite{Li:2010sv}.

Likewise, a scalar field with dual conformal weights $h_L=h_R=\frac{1}{2}+\beta$ reproduces the QNM spectrum associated with the inner photon shell in \eqref{eq:SpectrumMatching}.
The outer and inner photon shells are therefore shadow pairs.
Finally, the resonances of $\psi^{\rm out}(r)$ have the opposite sign for the imaginary part and grow rather than decay.
They are associated with the poles of the advanced Green's function, which are the poles in the upper half plane of the Feynman propagator \eqref{eq:FeynmanProp}.

This shows that the quantum Ruelle resonances of a CFT$_2$ can be matched to the spectrum of resonances of the warped AdS$_3$ spacetime \eqref{eq:WAdS3bis}, including its (anti-)QNMs.
Since the eikonal part of the QNM spectrum is governed by the photon shell, this strongly suggests that the dual CFT$_2$ encodes the photon ring and its Lyapunov exponents.

\section*{Acknowledgements}

We thank Shahar Hadar for fruitful discussions.
DK gratefully acknowledges support from 
the Center of Mathematical Sciences and Applications at Harvard University.
The work of AS is supported by DOE grant de-sc/0007870.
AL gratefully acknowledges Will and Kacie Snellings for their generous support and also thanks St\'ephane Detournay for useful conversations.

\appendix

\section{Anti-quasinormal mode spectrum}
\label{app:AQNM}

In this appendix, we collect formulas for anti-quasinormal modes that decay exponentially in the past. 
In section~\ref{subsec:Spectrum}, we considered the resonances of the warped AdS$_3$ metric \eqref{eq:WAdS3}, which occur when the leading boundary terms of the ingoing modes \eqref{eq:BoundaryIngoingModes} or outgoing modes \eqref{eq:BoundaryOutgoingModes} vanish.
We found that the resonant frequencies of the ``in'' modes correspond to the QNM spectrum \eqref{eq:QNM}.
The resonant frequencies of the ``out'' modes  are of the form $\frac{1}{2}\pm\beta+i\hat{\omega}=-n$ for some integer $n\in\mathbb{N}$, which results in the spectrum
\begin{align}
    \label{eq:AntiQNMspectrum}
	\omega_{mn\pm}=\mp2\pi T_H m\sqrt{\pa{1-\frac{1}{\Lambda^2}}-\frac{1}{4m^2}}+i\pa{n+\frac{1}{2}}2\pi T_H\;,\qquad
	m\in\mathbb{Z}\;,\quad
	n\in\mathbb{N}\;.
\end{align}
Since $\im\omega_{mn\pm}>0$, the associated modes decay exponentially in the past so $\omega_{mn\pm}$ is the spectrum of ``anti-QNMs.''

Section~\ref{subsec:ConformalSymmetry} reworked this calculation in the conformal coordinate system \eqref{eq:SimpleCoordinates}.
The resonances of the ``in'' modes, which make \eqref{eq:IngoingModesBoundary} vanish, correspond to the  QNMs \eqref{eq:QNMx} with frequencies \eqref{eq:FrequenciesQNM}.
The ``out'' mode resonances occur when $\frac{1}{2}\pm\beta+i\hat{\omega}=-n$, resulting in the spectrum \eqref{eq:AntiQNMspectrum} of anti-QNM frequencies.
The anti-QNMs are thus
\begin{align}
    \tilde{\Phi}_{mn\pm}(T,x,\phi)&=e^{-i\hat{\omega}T+im\phi-imx}\pa{\sinh{x}}^{-i\hat{\omega}}{_2F_1}\br{\frac{1}{2}+\beta+i\hat{\omega},\frac{1}{2}-\beta+i\hat{\omega};1+i\pa{\hat{\omega}+m};\frac{1}{1-e^{2x}}}\;,\\
	\hat{\omega}=\hat{\omega}_{mn\pm}&=\pm i\beta+i\pa{n+\frac{1}{2}}\;.
\end{align}
In the eikonal regime $\ab{m}\gg1$, it follows from \eqref{eq:EikonalBeta} that the fundamental $n=0$ mode takes the approximate form
\begin{align}
	\tilde{\Phi}_{m0\pm}(T,x,\phi)&\stackrel{\ab{m}\gg1}{\approx}e^{+\pa{\frac{1}{2}-im\hat{\Omega}_\pm}T+im\phi-imx}\pa{\sinh{x}}^{\frac{1}{2}-im\hat{\Omega}_\pm}\;.
\end{align}
From now on, we focus on the outer bound orbit at $\tilde{x}=\tilde{x}_+$ with angular velocity $\hat{\Omega}=\hat{\Omega}_+$, which controls the $+$ branch $\omega_{mn}=\omega_{mn+}$ of the anti-QNM spectrum.
Suppressing the $+$ subscript, we obtain
\begin{align}
	\label{eq:EikonalAQNM}
	\tilde{\Phi}_{m0}(T,x,\phi)&\stackrel{\ab{m}\gg1}{\approx}e^{+\pa{\frac{1}{2}-im\hat{\Omega}}T+im\phi-imx}\pa{\sinh{x}}^{\frac{1}{2}-im\hat{\Omega}}\;,
\end{align}
which is the analogue of \eqref{eq:EikonalQNM} for anti-QNMs.

Independent of the eikonal approximation, one can verify that the $n=0$ fundamental anti-QNM is lowest-weight and that its $n>0$ overtones are $\slr$ descendants.
Indeed, the modes
\begin{align}
    \label{eq:LWQNM}
	\Phi_{m0}^{\rm LW}(T,x,\phi)&=e^{+\hat{h}T+im\phi-imx}\pa{\sinh{x}}^{\hat{h}}\;,\qquad
	\hat{h}=\frac{1}{2}+\beta
	\stackrel{\ab{m}\gg1}{\approx}\frac{1}{2}-im\hat{\Omega}\;,
\end{align}
satisfy
\begin{align}
	\mathcal{L}_{\hat{L}_-}\Phi_{m0}^{\rm LW}=0\;,\qquad
	\mathcal{L}_{\hat{L}_0}\Phi_{m0}^{\rm LW}=-\hat{h}\Phi_{m0}^{\rm LW}\;,\qquad
	\mathcal{L}_{W_0}\Phi_{m0}^{\rm LW}=im\Phi_{m0}^{\rm LW}\;,
\end{align}
and their $\slr$ descendants obey
\begin{align}
	\Phi_{mn}^{\rm LW}=\mathcal{L}_{\hat{L}_+}^n\Phi_{m0}^{\rm LW}\;,\qquad
	\mathcal{L}_{\hat{L}_0}\Phi_{mn}^{\rm LW}=-\pa{\hat{h}+n}\Phi_{mn}^{\rm LW}\;.
\end{align}
These $\slr$ towers of modes reproduce the exponentially growing anti-QNM overtones up to normalization:
\begin{align}
	\Phi_{mn}^{\rm LW}=(-2)^n\br{\prod_{k=0}^{n-1}\hat{h}+k-im}\tilde{\Phi}_{mn}\;.
\end{align}
The exact value of the relative amplitude of successive overtones is 
\begin{align}
	\tilde{u}_m=\frac{\Phi_{m1}^{\rm LW}(T,x,\phi)}{\Phi_{m0}^{\rm LW}(T,x,\phi)}
	=-2e^{+T}\pa{\hat{h}\cosh{x}-im\sinh{x}}\;.
\end{align}
In the eikonal limit $\ab{m}\gg1$, we can approximate $\hat{h}\approx-im\hat{\Omega}=-\frac{im}{\hat{\lambda}}$, with $\hat{\lambda}=\hat{\lambda}_+$, simplifying this ratio to
\begin{align}
    \label{eq:EikonalRatioAQNM}
	\tilde{u}_m\approx2im\hat{\Omega}e^{+T}\pa{\cosh{x}+\hat{\lambda}\sinh{x}}
	\sim e^{+T}\pa{\cosh{x}+\hat{\lambda}\sinh{x}}\;.
 \end{align}

In section~\ref{subsec:NearRing}, we compared the exact QNMs to their near-ring approximations near the (outer) bound orbit $x\approx\tilde{x}$.
We now do the same for the exact anti-QNMs.
Letting $\dt x=x-\tilde{x}$ and using $\hat{h}\stackrel{\ab{m}\gg1}{\approx}-im\hat{\Omega}$,
\begin{align}
    e^{-imx}\pa{\sinh{x}}^{\hat{h}}=e^{-imx}e^{\hat{h}\log{\sinh{x}}}
    \stackrel{\ab{m}\gg1}{\approx}e^{-imx-im\hat{\Omega}\log{\sinh{x}}}
    \stackrel{\ab{\dt x}\ll1}{\approx}\tilde{c}_0e^{+\frac{i}{2}\frac{m\hat{\Omega}}{\Lambda^2-1}\dt x^2}.
\end{align}
Near the bound orbit, the exact anti-QNM wavefunctions \eqref{eq:LWQNM} with large $m$ therefore behave (up to a constant) as 
\begin{align}
    \label{eq:CompareExactAQNMNearRing}
	\Phi_{m0}^{\rm LW}(T,\dt x,\phi)\sim e^{+\frac{1}{2}T-im\hat{\Omega}\pa{T-\frac{1}{2}k\dt x^2}+im\phi}\;,\qquad
	k\equiv\frac{1}{\Lambda^2-1}\;.
\end{align}

In section~\ref{subsec:GeometricOptics}, we used the the geometric optics approximation to construct the eikonal QNMs using null congruences of unstable homoclinic rays.
Similarly, in this approximation, the eikonal anti-QNMs correspond to null congruences of stable homoclinic rays, with momentum and phase (up to an irrelevant phase shift by an integration constant $C$)
\begin{align}
    \hat{p}_\mu=\pd_\mu\hat{S}\;,\qquad
	\hat{S}&=\hat{\omega}\br{-T+\hat{\lambda}\phi-\int_{\tilde{x}}^x\pa{\frac{1}{\tanh{x'}}+\hat{\lambda}}\ed x'}
	=\hat{\omega}\br{-T+\hat{\lambda}\phi-\hat{\lambda}x-\log\pa{\sinh{x}}+C}\;.
\end{align}
These geodesics asymptote to the peak of their radial potential $V(x)$ and form congruences with precisely the right behavior to approximate wave solutions obeying anti-QNM boundary conditions (namely, purely outgoing at the horizon and purely ingoing at the boundary).
The expansion of these congruences is
\begin{align}
	\theta=\nabla^2\hat{S}
	=\frac{\hat{\omega}}{\ell^2}\;.
\end{align}
The subleading equation \eqref{eq:GeometricAmplitude} for the wave amplitude $\tilde{A}(T,x)$ takes the explicit form
\begin{align}
	\br{\pa{\frac{1}{\tanh{x}}+\hat{\lambda}}\pd_x-\pa{1+\frac{\hat{\lambda}}{\tanh{x}}}\pd_T-\frac{1}{2\sinh^2{x}}}\tilde{A}(T,x)=0\;,
\end{align}
which admits a minimal solution $\tilde{A}_0(T,x)=e^{+\frac{1}{2}T}\sqrt{\sinh{x}}$ that generates an infinite family of solutions of the form
\begin{align}
    \label{eq:OvertoneRatioAQNM}
	\tilde{A}_n(T,x)=\tilde{u}^n(T,x)\tilde{A}_0(T,x)\;,\qquad
	\tilde{u}(T,x)=e^{+T}\pa{\cosh{x}+\hat{\lambda}\sinh{x}}\;.
\end{align}
Recalling that $\hat{\omega}=m\hat{\Omega}$, the eikonal approximation to the anti-QNMs is
\begin{align}
	\label{eq:ApproxAQNM}
	\tilde{\Phi}_{mn}(T,x,\phi)\approx \tilde{A}_n(T,x)e^{i\hat{S}(T,x,\phi)}
	=e^{+nT}\pa{\cosh{x}+\hat{\lambda}\sinh{x}}^ne^{+\pa{\frac{1}{2}-im\hat{\Omega}}T+im\phi-imx}\pa{\sinh{x}}^{\frac{1}{2}-im\hat{\Omega}}\;.
\end{align}
In particular, setting $n=0$ in this formula results in
\begin{align}
	\tilde{\Phi}_{m0}(T,x,\phi)\approx \tilde{A}_0(T,x)e^{i\hat{S}(T,x,\phi)}
	=e^{+\pa{\frac{1}{2}- im\hat{\Omega}}T+im\phi-imx}\pa{\sinh{x}}^{\frac{1}{2}-im\hat{\Omega}}\;,
\end{align}
which precisely reproduces the eikonal limit \eqref{eq:EikonalAQNM} of the fundamental anti-QNM.
Likewise, the geodesic approximation \eqref{eq:ApproxAQNM} predicts that successive overtones differ by the ratio $\tilde{u}(T,x)$ given in \eqref{eq:OvertoneRatioAQNM}, which also matches the eikonal limit of the exact ratio given in \eqref{eq:EikonalRatioAQNM} for anti-QNMs.
Note that to first order in the near-ring region
\begin{align}
    \tilde{u}(T,\tilde{x}+\dt x)\approx-\frac{e^{+T}\dt x}{\sqrt{\Lambda^2-1}}\;,
\end{align}
which is indeed a constant along the stable homoclinic orbits according to \eqref{eq:GeodesicDeviation}--\eqref{eq:LyapunovExponent}.

On a stable homoclinic trajectory with critical momentum $\hat{p}$, the Hamiltonian functions \eqref{eq:Functions} in the near-ring region $x=\tilde{x}+\dt x$ take the approximate form 
\begin{align}
    \hat{L}_-(\hat{p})=0\;,\qquad
    \hat{L}_+(\hat{p})=2\hat{\omega}e^{+T}\pa{\cosh{x}+\hat{\lambda}\sinh{x}}
    \propto\tilde{u}(T,x)\;.
\end{align}
This is equivalent to the statement that the phase function $\Psi_0=e^{i\hat{S}}$ is lowest-weight with respect to $\slr_{\rm ISO}$, and we could equivalently express \eqref{eq:OvertoneRatioAQNM} as 
\begin{align}
    \tilde{A}_n(T,x)\propto\hat{L}_+^n(T,x,\hat{p})\tilde{A}_0(T)\;.  
\end{align}
This equation states that to generate higher-overtone amplitudes based on a homoclinic congruence, one simply multiplies the seed amplitude by the Hamiltonian function $\hat{L}_+(x,\hat{p})$.
This is the geometric optics analogue of the exact statement that the higher-overtone anti-QNM wavefunctions are conformal descendants
\begin{align}
    \tilde{\Phi}_{mn}\propto\mathcal{L}_{\hat{L}_+}^n\tilde{\Phi}_{m0}\;. 
\end{align}

\section{Geodesic phase space reduction and symmetries}
\label{app:Reduction}

Geodesic motion is usually defined in the extended phase space (EPS) $(t,x^i,p_t,p_i)$ corresponding to \textit{spacetime} rather than the reduced phase space (RPS) $(x^i,p_i)$ corresponding to space.
The RPS formulation is not covariant: it requires a splitting of spacetime into spatial slices, and the reduced Hamiltonian $H$ then generates time evolution with respect to the chosen coordinate time.
By contrast, the EPS formulation is covariant, and the extended Hamiltonian $\mathcal{H}$ can produce geodesics with any parameterization; typically, one studies the affinely parameterized geodesics generated by
\begin{align}
	\mathcal{H}(t,x^i,p_t,p_i)=\frac{1}{2}g^{\mu\nu}p_\mu p_\nu\;,
\end{align}
with the canonical symplectic form $\Omega=\ed p_\mu\wedge\ed x^\mu$.
A downside of this EPS system is that it is subject to a constraint
\begin{align}
	\mathcal{H}=-\frac{1}{2}\mu^2\;,
\end{align}
where $\mu^2=-p_\mu p^\mu$ is the invariant mass of the particle on the geodesic.
Solving this dynamical constraint for
\begin{align}
	H(x^i,p_i)=-p_t
\end{align}
reproduces the Hamiltonian of the RPS formulation, with canonical symplectic form $\tilde{\Omega}=\ed p_i\wedge\ed x^i$.
This procedure---whereby a dynamical constraint is eliminated in favor of a lower-dimensional formulation---is known as \textit{phase space reduction}.
The reduction from EPS to RPS at time $t_0$ is achieved by sending $(t,p_t)\to(t_0,-H)$, such that $\tilde{\Omega}\to\Omega$.

\subsection*{Global and local conformal symmetry of phase space}
\label{app:Flows}

In section~\ref{sec:Symmetries}, we defined an action of the global conformal group $\slr_{\rm PR}$ on the (reduced) phase space of null geodesics in the geometry \eqref{eq:SimpleWAdS3}.
The same construction was also carried out for Schwarzschild and Kerr in \cite{Hadar2022}.
In all these cases, the $\slr_{\rm PR}$ dilations map successive subring images to one another.
This construction is general and only requires the existence of canonical coordinates $(T,H,\ldots)$ in terms of which $\slr_{\rm PR}$ generators are defined as
\begin{align}
    \label{eq:SimpleSL2R}
	H_n=(-T)^{1-n}H\;,\qquad
	 n\in\cu{-1,0,1}\;.
\end{align}
Extending this definition to arbitrary $n\in\mathbb{Z}$ yields generators of a full, infinite-dimensional Witt algebra
\begin{align}
	\cu{H_m,H_n}=(m-n)H_{m+n}\;,\qquad
	\forall m,n\in\mathbb{Z}\;.
\end{align}
Since $\cu{T,H_n}=(-T)^{1-n}$ and $\{H,H_n\}=(1-n)H(-T)^{-n}$, the function $H_n$ (with $n\neq0$) generates the RPS flow
\begin{align}
	T(s)=-\br{n(c_1-s)}^{1/n}\;,\qquad
	H(s)=\frac{c_2(c_1-s)}{T(s)}\;,
\end{align}
for some constants $c_1$ and $c_2$.
For $\ab{n}\ge2$, this flow becomes manifestly complex and multi-valued past $s=c_1$ due to the branch cut in the $n^\text{th}$ root.
Hence, the extension to an infinite-dimensional conformal symmetry is only locally (but not globally) defined, as in CFT$_2$.
However, the $\slr_{\rm PR}$ subalgebra generates globally well-defined flows:
\begin{align}
	&e^{sL_+}\;:
	&&T(s)=s+T(0)\;,
	&&H(s)=H(0)\;,\\
	\label{eq:DilationAction}
	&e^{sL_0}\;:
	&&T(s)=e^{-s}T(0)\;,
	&&H(s)=e^sH(0)\;,\\
	&e^{sL_-}\;:
	&&T(s)=\frac{T(0)}{1-sT(0)}\;,
	&&H(s)=H(0)\br{1-sT(0)}^2\;.
\end{align}
A general finite $\slr_{\rm PR}$ transformation acts as
\begin{align}
	T\to\frac{aT+b}{cT+d}\;,\qquad
	H\to\pa{cT+d}^2H\;,\qquad
	ad-bc=1\;.
\end{align}
This is a symplectomorphism since $\ed H\wedge\ed T$ is left invariant.
This is because $\slr_{\rm PR}\cong\mathsf{Sp}(2,\mathbb{R})$: the symplectic group of canonical transformations on a two-dimensional phase space is the global conformal group.

\subsection*{Ambiguity in special conformal transformations and Casimir of \texorpdfstring{$\slr_{\rm PR}$}{SL(2,R)}}

The $\slr_{\rm PR}$ group generated by the functions in \eqref{eq:SimpleSL2R} has two possible ambiguities.
The first is trivial and consists of shifts in $H$ by functions that commute with both $T$ and $H$, thus leaving the $\slr_{\rm PR}$ algebra invariant.
In fact, such a shift is present in the $\slr_{\rm PR}$ generators \eqref{eq:SL2RPR} for the warped spacetime \eqref{eq:SimpleWAdS3}, as well as for their analogues in Schwarzschild and Kerr \cite{Hadar2022}. These constructions involve $\hat{H}=H-\tilde{H}$ rather than the geodesic Hamiltonian $H$ by itself: a shift by the critical Hamiltonian $\tilde{H}$ is crucial to ensure that $\slr_{\rm PR}$ transformations scale arbitrary geodesics into the critical bound orbit, leaving the photon shell as the only $\slr_{\rm PR}$-invariant phase space submanifold.

The second ambiguity is more interesting: we note that one can also shift the special conformal generator $H_-$ by a term of the form $\mathcal{C}/H_+$ without affecting the $\slr_{\rm PR}$ commutation relations,
\begin{align}
    \label{eq:CasimirChoice}
	H_+=\hat{H}\;,\qquad
	H_0=-TH_+\;,\qquad
	H_-=T^2H_++\frac{\mathcal{C}}{H_+}\;,
\end{align}
provided that $\mathcal{C}$ is a conserved quantity along geodesics that commutes with both $T$ and $H$, and therefore with all of $\slr_{\rm PR}$.
In fact, we recognize this quantity $\mathcal{C}$ as the Casimir of $\slr_{\rm PR}$,
\begin{align}
    \mathcal{C}=-H_0^2+\frac{H_+H_-+H_-H_+}{2}\;.
\end{align}
Since we only used the dilation generator $H_0$ to relate successive photon subrings, the particular form of the special conformal transformation generator $H_-$ was not important to us, and we set it to zero in \eqref{eq:SL2RPR}, as we did in Schwarzschild and Kerr \cite{Hadar2022}.
However, these spacetimes are all axisymmetric, and hence $\mathcal{C}$ could be any function of the angular momentum $p_\phi$ that is conserved along null geodesics (as well as of the Carter constant in Kerr).
As we showed in section~\ref{sec:Symmetries}, in self-dual warped AdS$_3$, this Casimir must take the specific form $\mathcal{C}=-\pa{1-\Lambda^{-2}}p_\phi^2$ in order for this construction to recover the spacetime isometry group.sc

\subsection*{Constructing the time coordinate \texorpdfstring{$T$}{T}}

In order to define $\slr_{\rm PR}$ as in \eqref{eq:SimpleSL2R}, it is necessary to first construct the time coordinate $T(x^i,p_i)$ on RPS that is canonically conjugate to the Hamiltonian $H$, i.e., that satisfies $\cu{T,H}_{\rm RPS}=1$.

In section~\ref{sec:Symmetries}, we used the canonical transformation \eqref{eq:ActionAngle} to reduce the geodesic equation in the geometry \eqref{eq:SimpleWAdS3} to a problem of quadratures, resulting in the integral expression \eqref{eq:TimeLapse} for $T(x,H,L)$.
A more direct approach takes advantage of the EPS formulation, where the time coordinate obeys $\dot{t}=\cu{t,\mathcal{H}}_{\rm EPS}$.
Thus, one may construct a function $\mathcal{T}(x^i,p_t,p_i)$ that also keeps track of the time elapsed along a geodesic by demanding that $\dot{t}=\cu{\mathcal{T},\mathcal{H}}_{\rm EPS}$, or
\begin{align}
    \label{eq:TimeFunction}
    \cu{\mathcal{T},\mathcal{H}}_{\rm EPS}=\pd_{p_t}\mathcal{H}\;.
\end{align}
One can then directly recover the time function $T$ on RPS as $T(x^i,p_i)=\mathcal{T}(x^i,p_t,p_i)|_{p_t=-H(x^i,p_i)}$.

For example, null geodesic motion in the warped AdS$_3$ spacetime \eqref{eq:WAdS3} is generated by the EPS Hamiltonian
\begin{align}
    \label{eq:HamiltonianEPS}
    \mathcal{H}=\frac{1}{2}g^{\mu\nu}p_\mu p_\nu
    =\frac{1}{2\ell^2}\br{r\pa{r+4\pi T_H}p_r^2-\frac{\br{p_t-\pa{r+2\pi T_H}p_\phi}^2}{r\pa{r+4\pi T_H}}+\frac{p_\phi^2}{\Lambda^2}}
    \equiv0\;,
\end{align}
while the RPS Hamiltonian $H=-p_t$ is obtained by solving $\mathcal{H}=0$, which results in
\begin{align}
    \label{eq:HamiltonianRPS}
    H=-\pa{r+2\pi T_H}p_\phi-r\pa{r+4\pi T_H}\sqrt{p_r^2+\frac{p_\phi^2}{\Lambda^2r\pa{r+4\pi T_H}}}\;.
\end{align}
The time function $\mathcal{T}(r,\phi,p_t,p_r,p_\phi)$ is defined by the EPS condition \eqref{eq:TimeFunction}, which yields
\begin{align}
    \label{eq:ExtendedTime}
    \mathcal{T}=\frac{1}{4\pi T_H}\log\br{\frac{\pa{r+2\pi T_H}p_t-2\pi T_H\br{2\pi T_Hp_\phi-r\pa{r+4\pi T_H}p_r}}{\pa{r+2\pi T_H}p_t-2\pi T_H\br{2\pi T_Hp_\phi+r\pa{r+4\pi T_H}p_r}}}-\mathcal{T}_0\;,
\end{align}
where $\mathcal{T}_0$ is an arbitrary integration constant.
Plugging in $p_t=-H(r,p_r,p_\phi)$ yields a function $T(r,p_r,p_\phi)$ such that $\cu{T,H}_{\rm RPS}=1$, as desired.
It is written explicitly in \eqref{eq:SolveTime} below.

In the zero-temperature limit $T_H\to0$, the EPS and RPS Hamiltonians \eqref{eq:HamiltonianEPS} and \eqref{eq:HamiltonianRPS} reduce to \eqref{eq:ZeroHamiltonianEPS} and \eqref{eq:ZeroHamiltonianRPS}, respectively, while the EPS and RPS times \eqref{eq:ExtendedTime} and \eqref{eq:SolveTime} become singular, with \eqref{eq:ZeroTime} replacing the latter.

\section{\texorpdfstring{$\slr_{\rm QN}$ and $\slr_{\rm ISO}^2$}{Quasinormal SL(2,R) and Isometry SL(2,R)}}

Here, we focus on the warped AdS$_3$ metric \eqref{eq:WAdS3} written in the coordinates $(t,r,\phi)$ that make the dependence on the black hole temperature $T_H$ explicit.
This geometry admits an isometry group $\slr_{\rm ISO}$ generated by vectors
\begin{align}
    \label{eq:AnotherSL2R}
    \hat{L}_0=-\frac{1}{2\pi T_H}\pd_t\;,\qquad
    \hat{L}_\pm=e^{\pm2\pi T_Ht}\br{-\frac{r+2\pi T_H}{2\pi T_H\sqrt{r\pa{r+4\pi T_H}}}\pd_t\pm\sqrt{r\pa{r+4\pi T_H}}\pd_r+\frac{2\pi T_H}{\sqrt{r\pa{r+4\pi T_H}}}\pd_\phi}\;,
\end{align}
which are of course equivalent to \eqref{eq:Isometries} under the coordinate transformation \eqref{eq:SimpleCoordinates}.
These Killing vector fields define functions $\hat{L}_n=-\hat{L}_n^\mu p_\mu$ that generate the action of $\slr_{\rm ISO}$ on the six-dimensional phase space $(t,r,\phi,p_t,p_r,p_\phi)$ with canonical symplectic form.
Null geodesic motion in that phase space is generated by the constrained Hamiltonian \eqref{eq:HamiltonianEPS}, and the fact that the vector fields $\hat{L}_n^\mu\pd_\mu$ are Killing ensures that the functions $\hat{L}_n$ are conserved along geodesics:
\begin{align}
    \cu{\hat{L}_n,\mathcal{H}}=0\;,\qquad
	\forall n\in\cu{-1,0,1}\;.
\end{align}
We note however that these conserved charges do not commute with each other, as they obey the $\slr_{\rm ISO}$ algebra.
The Casimir of this $\slr_{\rm ISO}$ is
\begin{align}
    \mathcal{C}=-\hat{L}_0^2+\frac{\hat{L}_+\hat{L}_-+\hat{L}_-\hat{L}_+}{2}
    =-\pa{1-\frac{1}{\Lambda^2}}p_\phi^2\;.
\end{align}

In the four-dimensional phase space $(r,\phi,p_r,p_\phi)$, null geodesic motion is generated by the Hamiltonian $H=-p_t$ given in \eqref{eq:HamiltonianRPS}.
Its canonically conjugate variable with respect to the (four-dimensional) canonical symplectic form is the coordinate time $T(r,\phi,p_r,p_\phi)$ elapsed along geodesic flow,
\begin{align}
    \label{eq:Time}
    \cu{T,H}=1\;,
\end{align}
and, as explained below \eqref{eq:ExtendedTime}, it is given (up to an integration constant $\mathcal{T}_0$) by
\begin{align}
    \label{eq:SolveTime}
    T=\frac{1}{4\pi T_H}\log\br{\frac{\pa{r+2\pi T_H}H+2\pi T_H\br{2\pi T_Hp_\phi-r\pa{r+4\pi T_H}p_r}}{\pa{r+2\pi T_H}H+2\pi T_H\br{2\pi T_Hp_\phi+r\pa{r+4\pi T_H}p_r}}}-\mathcal{T}_0\;.
\end{align}

Under phase space reduction at time $t=t_0$, the functions $\hat{L}_n=-\hat{L}_n^\mu p_\mu$ defined from \eqref{eq:AnotherSL2R} reduce to
\begin{align}
    \label{eq:SGSL2R}
    \hat{\ell}_0=-\frac{H}{2\pi T_H}\;,\qquad
    \hat{\ell}_\pm=e^{\pm2\pi T_Ht_0}\frac{\sqrt{r\pa{r+4\pi T_H}}}{2\pi T_H}\br{p_\phi+\pa{r+2\pi T_H}\sqrt{p_r^2+\frac{p_\phi^2}{\Lambda^2r\pa{r+4\pi T_H}}}\mp2\pi T_Hp_r}\;,
\end{align}
where $t_0$ is just a constant and the factors of $e^{\pm2\pi T_Ht_0}$ could be removed by a dilation $\hat{\ell}_\pm\to\lambda^{\pm1}\hat{\ell}_\pm$ with $\lambda=e^{-2\pi T_Ht_0}$, which leaves the $\slr$ algebra invariant (as $\slr$ is its own automorphism group).
We note that the functions $\hat{\ell}_\pm$ are \textit{no longer} conserved in four-dimensional phase space, since they do not commute with the Hamiltonian $H$ and therefore change the energy (or photon color) of a null geodesic.
We also note that after setting $\mathcal{T}_0=t_0$,
\begin{align}
    T=\frac{1}{4\pi T_H}\log\pa{\frac{\hat{\ell}_-}{\hat{\ell}_+}}\;,\qquad
    \mathcal{C}\equiv-\hat{\ell}_0^2+\frac{\hat{\ell}_+\hat{\ell}_-+\hat{\ell}_-\hat{\ell}_+}{2}
    =-\pa{1-\frac{1}{\Lambda^2}}p_\phi^2\;.
\end{align}
Inverting these relations allows us to rewrite \eqref{eq:SGSL2R} as
\begin{align}
    \hat{\ell}_0=-\frac{H}{2\pi T_H}\;,\qquad
    \hat{\ell}_\pm=e^{\mp2\pi T_HT}\sqrt{\hat{\ell}_0^2+\mathcal{C}}\;,\qquad
    \mathcal{C}=-\pa{1-\frac{1}{\Lambda^2}}p_\phi^2\,,
\end{align}
which makes it clear that the $\hat{\ell}_\pm$ vanish when and only when they are evaluated on the critical bound orbit $p_\phi=\tilde{\lambda}H$, with $\tilde{\lambda}$ given in \eqref{eq:CriticalImpact}.
In this form, the fact that these functions obey the $\slr_{\rm ISO}$ commutation relations follows immediately from the commutation relation \eqref{eq:Time}.

To compare this $\slr_{\rm ISO}$ with the $\slr_{\rm QN}$ introduced in \eqref{eq:SL2RQN}, we must return to coordinates \eqref{eq:SimpleCoordinates}, which 
results in\footnote{If not for the presence of a nonzero Casimir $\mathcal{C}$, we would have $\hat{\ell}_\pm=e^{\pm T}H$ and these functions would be Susskind-Glogower operators.}
\begin{align}
    \hat{\ell}_0=-H
    =\frac{p_\phi}{\tanh{x}}+\sqrt{p_x^2+\frac{p_\phi^2}{\Lambda^2\sinh^2{x}}}\;,\qquad
    \hat{\ell}_\pm=e^{\mp T}\sqrt{H^2+\mathcal{C}}
    =-H\cosh{x}-\pa{p_\phi\mp p_x}\sinh{x}\;.
\end{align}
The phase space coordinates are now $(x,\phi,p_x,p_\phi)$.
To scale into the near-ring region, we change coordinates to
\begin{align}
    x=\tilde{x}+\epsilon\dt x\;,\qquad
    p_x=\epsilon P\;,
\end{align}
which is a canonical transformation provided that we set the dummy parameter $\epsilon$ to unity after expanding near the ring in small $\epsilon$.
We thus find that to leading order near the ring,
\begin{align}
    \hat{\ell}_\pm\approx\sqrt{\Lambda^2-1}\pa{\pm P+kH\dt x}\;,
\end{align}
where $k=\pa{\Lambda^2-1}^{-1}$ as defined in \eqref{eq:CompareExactNearRing}.
To make contact with the radial wave equation from section~\ref{subsec:NearRing}, we quantize these operators by letting $P=-i\pd_{\dt x}$, in which case they act as
\begin{align}
    \hat{\ell}_\pm\approx i\sqrt{\Lambda^2-1}\pa{\mp\pd_{\dt x}-ikH\dt x}
    \propto a_\pm\;,
\end{align}
provided that we identify the energy as $H=\hat{\omega}_R$.
Since the generators $L_\pm$ of $\slr_{\rm QN}$ are obtained by squaring the near-ring generators $a_\pm$, we conclude that $\slr_{\rm QN}$ is related to the square of the near-ring isometry group $\slr_{\rm ISO}$.
This conclusion is consistent with \eqref{eq:Squaring}.

\bibliography{kls}

\providecommand{\href}[2]{#2}\begingroup\raggedright\begin{thebibliography}{10}

\bibitem{Johnson2020}
M.~D. {Johnson}, A.~{Lupsasca}, A.~{Strominger}, G.~N. {Wong}, S.~{Hadar},
  D.~{Kapec}, R.~{Narayan}, A.~{Chael}, C.~F. {Gammie}, P.~{Galison}, D.~C.~M.
  {Palumbo}, S.~S. {Doeleman}, L.~{Blackburn}, M.~{Wielgus}, D.~W. {Pesce},
  J.~R. {Farah}, and J.~M. {Moran}, ``{Universal interferometric signatures of
  a black hole's photon ring},''
  \href{http://dx.doi.org/10.1126/sciadv.aaz1310}{{\em Science Advances}
  {\bfseries 6} no.~12, (Mar., 2020) eaaz1310},
  \href{http://arxiv.org/abs/1907.04329}{{\ttfamily arXiv:1907.04329
  [astro-ph.IM]}}.

\bibitem{Abbott2016}
{LIGO Scientific Collaboration} and {Virgo Collaboration}, ``{Observation of
  Gravitational Waves from a Binary Black Hole Merger},''
  \href{http://dx.doi.org/10.1103/PhysRevLett.116.061102}{{\em \prl} {\bfseries
  116} no.~6, (Feb., 2016) 061102},
  \href{http://arxiv.org/abs/1602.03837}{{\ttfamily arXiv:1602.03837 [gr-qc]}}.

\bibitem{Hadar2022}
S.~Hadar, D.~Kapec, A.~Lupsasca, and A.~Strominger, ``{Holography of the photon
  ring},'' \href{http://dx.doi.org/10.1088/1361-6382/ac8d43}{{\em Class. Quant.
  Grav.} {\bfseries 39} no.~21, (2022) 215001},
  \href{http://arxiv.org/abs/2205.05064}{{\ttfamily arXiv:2205.05064 [gr-qc]}}.

\bibitem{Israel2005}
D.~{Isra{\"e}l}, C.~{Kounnas}, D.~{Orlando}, and P.~{Marios Petropoulos},
  ``{Electric/magnetic deformations of S3 and AdS3, and geometric cosets},''
  \href{http://dx.doi.org/10.1002/prop.200410190}{{\em Fortschritte der Physik}
  {\bfseries 53} no.~1, (Jan., 2005) 73--104},
  \href{http://arxiv.org/abs/hep-th/0405213}{{\ttfamily arXiv:hep-th/0405213
  [hep-th]}}.

\bibitem{Detournay2011}
S.~{Detournay}, D.~{Isra{\"e}l}, J.~M. {Lapan}, and M.~{Romo}, ``{String theory
  on warped AdS$_{3}$ and Virasoro resonances},''
  \href{http://dx.doi.org/10.1007/JHEP01(2011)030}{{\em \jhep} {\bfseries 2011}
  (Jan., 2011) 30}, \href{http://arxiv.org/abs/1007.2781}{{\ttfamily
  arXiv:1007.2781 [hep-th]}}.

\bibitem{Azeyanagi2013}
T.~{Azeyanagi}, D.~M. {Hofman}, W.~{Song}, and A.~{Strominger}, ``{The spectrum
  of strings on warped AdS$_{3}\times\mathrm{S}^{3}$},''
  \href{http://dx.doi.org/10.1007/JHEP04(2013)078}{{\em \jhep} {\bfseries 2013}
  (Apr., 2013) 78}, \href{http://arxiv.org/abs/1207.5050}{{\ttfamily
  arXiv:1207.5050 [hep-th]}}.

\bibitem{Hofman2011}
D.~M. {Hofman} and A.~{Strominger}, ``{Chiral Scale and Conformal Invariance in
  2D Quantum Field Theory},''
  \href{http://dx.doi.org/10.1103/PhysRevLett.107.161601}{{\em \prl} {\bfseries
  107} no.~16, (Oct., 2011) 161601},
  \href{http://arxiv.org/abs/1107.2917}{{\ttfamily arXiv:1107.2917 [hep-th]}}.

\bibitem{Detournay2012}
S.~{Detournay}, T.~{Hartman}, and D.~M. {Hofman}, ``{Warped conformal field
  theory},'' \href{http://dx.doi.org/10.1103/PhysRevD.86.124018}{{\em \prd}
  {\bfseries 86} no.~12, (Dec., 2012) 124018},
  \href{http://arxiv.org/abs/1210.0539}{{\ttfamily arXiv:1210.0539 [hep-th]}}.

\bibitem{Bzowski2019}
A.~{Bzowski} and M.~{Guica}, ``{The holographic interpretation of
  $J\bar{T}$-deformed CFTs},''
  \href{http://dx.doi.org/10.1007/JHEP01(2019)198}{{\em Journal of High Energy
  Physics} {\bfseries 2019} no.~1, (Jan., 2019) 198},
  \href{http://arxiv.org/abs/1803.09753}{{\ttfamily arXiv:1803.09753
  [hep-th]}}.

\bibitem{Guica2018}
M.~{Guica}, ``{An integrable Lorentz-breaking deformation of two-dimensional
  CFTs},'' \href{http://dx.doi.org/10.21468/SciPostPhys.5.5.048}{{\em SciPost
  Physics} {\bfseries 5} no.~5, (Nov., 2018) 048},
  \href{http://arxiv.org/abs/1710.08415}{{\ttfamily arXiv:1710.08415
  [hep-th]}}.

\bibitem{Chakraborty2018}
S.~{Chakraborty}, A.~{Giveon}, and D.~{Kutasov}, ``{$J\bar{T}$ deformed
  CFT$_{2}$ and string theory},''
  \href{http://dx.doi.org/10.1007/JHEP10(2018)057}{{\em \jhep} {\bfseries 2018}
  no.~10, (Oct., 2018) 57}, \href{http://arxiv.org/abs/1806.09667}{{\ttfamily
  arXiv:1806.09667 [hep-th]}}.

\bibitem{Apolo2018}
L.~{Apolo} and W.~{Song}, ``{Strings on warped AdS$_{3}$ via $T\bar{J}$
  deformations},'' \href{http://dx.doi.org/10.1007/JHEP10(2018)165}{{\em \jhep}
  {\bfseries 2018} no.~10, (Oct., 2018) 165},
  \href{http://arxiv.org/abs/1806.10127}{{\ttfamily arXiv:1806.10127
  [hep-th]}}.

\bibitem{Chakraborty2019}
S.~{Chakraborty}, A.~{Giveon}, and D.~{Kutasov}, ``{$T\bar{T}$, $J\bar{T}$,
  $T\bar{J}$ and String Theory},''
  \href{http://dx.doi.org/10.1088/1751-8121/ab3710}{{\em Journal of Physics A}
  {\bfseries 52} no.~38, (2019) 384003},
  \href{http://arxiv.org/abs/1905.00051}{{\ttfamily arXiv:1905.00051
  [hep-th]}}.

\bibitem{Apolo2020}
L.~{Apolo} and W.~{Song}, ``{Heating up holography for single-trace $J\bar{T}$
  deformations},'' \href{http://dx.doi.org/10.1007/JHEP01(2020)141}{{\em \jhep}
  {\bfseries 2020} no.~1, (Jan., 2020) 141},
  \href{http://arxiv.org/abs/1907.03745}{{\ttfamily arXiv:1907.03745
  [hep-th]}}.

\bibitem{Guica2022}
M.~{Guica}, ``{A definition of primary operators in $J\bar{T}$-deformed
  CFTs},'' \href{http://dx.doi.org/10.21468/SciPostPhys.13.3.045}{{\em SciPost
  Physics} {\bfseries 13} no.~3, (Sept., 2022) 045},
  \href{http://arxiv.org/abs/2112.14736}{{\ttfamily arXiv:2112.14736
  [hep-th]}}.

\bibitem{Anninos2009}
W.~{Song}, D.~{Anninos}, W.~{Li}, M.~{Padi}, and A.~{Strominger}, ``{Warped
  AdS$_{3}$ black holes},''
  \href{http://dx.doi.org/10.1088/1126-6708/2009/03/130}{{\em \jhep} {\bfseries
  2009} no.~3, (Mar., 2009) 130},
  \href{http://arxiv.org/abs/0807.3040}{{\ttfamily arXiv:0807.3040 [hep-th]}}.

\bibitem{Chen2009}
B.~{Chen} and Z.-b. {Xu}, ``{Quasi-normal modes of warped black holes and
  warped AdS/CFT correspondence},''
  \href{http://dx.doi.org/10.1088/1126-6708/2009/11/091}{{\em \jhep} {\bfseries
  2009} no.~11, (Nov., 2009) 091},
  \href{http://arxiv.org/abs/0908.0057}{{\ttfamily arXiv:0908.0057 [hep-th]}}.

\bibitem{Chen2010a}
B.~{Chen}, B.~{Ning}, and Z.-B. {Xu}, ``{Real-time correlators in warped
  AdS/CFT correspondence},''
  \href{http://dx.doi.org/10.1007/JHEP02(2010)031}{{\em \jhep} {\bfseries 2010}
  (Feb., 2010) 31}, \href{http://arxiv.org/abs/0911.0167}{{\ttfamily
  arXiv:0911.0167 [hep-th]}}.

\bibitem{Chen2010b}
B.~{Chen} and J.~{Long}, ``{Hidden conformal symmetry and quasinormal modes},''
  \href{http://dx.doi.org/10.1103/PhysRevD.82.126013}{{\em \prd} {\bfseries 82}
  no.~12, (Dec., 2010) 126013},
  \href{http://arxiv.org/abs/1009.1010}{{\ttfamily arXiv:1009.1010 [hep-th]}}.

\bibitem{Song2012}
W.~{Song} and A.~{Strominger}, ``{Warped AdS$_{3}$/dipole-CFT duality},''
  \href{http://dx.doi.org/10.1007/JHEP05(2012)120}{{\em \jhep} {\bfseries 2012}
  (May, 2012) 120}, \href{http://arxiv.org/abs/1109.0544}{{\ttfamily
  arXiv:1109.0544 [hep-th]}}.

\bibitem{Moussa2003}
K.~A. {Moussa}, G.~{Cl{\'e}ment}, and C.~{Leygnac}, ``{The black holes of
  topologically massive gravity},''
  \href{http://dx.doi.org/10.1088/0264-9381/20/24/L01}{{\em Classical and
  Quantum Gravity} {\bfseries 20} no.~24, (Dec., 2003) L277--L283},
  \href{http://arxiv.org/abs/gr-qc/0303042}{{\ttfamily arXiv:gr-qc/0303042
  [gr-qc]}}.

\bibitem{Bouchareb2007}
A.~{Bouchareb} and G.~{Cl{\'e}ment}, ``{Black hole mass and angular momentum in
  topologically massive gravity},''
  \href{http://dx.doi.org/10.1088/0264-9381/24/22/018}{{\em Classical and
  Quantum Gravity} {\bfseries 24} no.~22, (Nov., 2007) 5581--5594},
  \href{http://arxiv.org/abs/0706.0263}{{\ttfamily arXiv:0706.0263 [gr-qc]}}.

\bibitem{Bredberg2010}
I.~{Bredberg}, T.~{Hartman}, W.~{Song}, and A.~{Strominger}, ``{Black hole
  superradiance from Kerr/CFT},''
  \href{http://dx.doi.org/10.1007/JHEP04(2010)019}{{\em \jhep} {\bfseries 2010}
  (Apr., 2010) 19}, \href{http://arxiv.org/abs/0907.3477}{{\ttfamily
  arXiv:0907.3477 [hep-th]}}.

\bibitem{Kapec2020}
D.~{Kapec} and A.~{Lupsasca}, ``{Particle motion near high-spin black holes},''
  \href{http://dx.doi.org/10.1088/1361-6382/ab519e}{{\em Classical and Quantum
  Gravity} {\bfseries 37} no.~1, (Jan., 2020) 015006},
  \href{http://arxiv.org/abs/1905.11406}{{\ttfamily arXiv:1905.11406
  [hep-th]}}.

\bibitem{Birmingham2002}
D.~{Birmingham}, I.~{Sachs}, and S.~N. {Solodukhin}, ``{Conformal Field Theory
  Interpretation of Black Hole Quasinormal Modes},''
  \href{http://dx.doi.org/10.1103/PhysRevLett.88.151301}{{\em \prl} {\bfseries
  88} no.~15, (Apr., 2002) 151301},
  \href{http://arxiv.org/abs/hep-th/0112055}{{\ttfamily arXiv:hep-th/0112055
  [hep-th]}}.

\bibitem{Birmingham2003}
D.~{Birmingham}, I.~{Sachs}, and S.~N. {Solodukhin}, ``{Relaxation in conformal
  field theory, Hawking-Page transition, and quasinormal or normal modes},''
  \href{http://dx.doi.org/10.1103/PhysRevD.67.104026}{{\em \prd} {\bfseries 67}
  no.~10, (May, 2003) 104026},
  \href{http://arxiv.org/abs/hep-th/0212308}{{\ttfamily arXiv:hep-th/0212308
  [hep-th]}}.

\bibitem{Song2018}
W.~{Song} and J.~{Xu}, ``{Correlation functions of warped CFT},''
  \href{http://dx.doi.org/10.1007/JHEP04(2018)067}{{\em \jhep} {\bfseries 2018}
  no.~4, (Apr., 2018) 67}, \href{http://arxiv.org/abs/1706.07621}{{\ttfamily
  arXiv:1706.07621 [hep-th]}}.

\bibitem{Guica2009}
M.~{Guica}, T.~{Hartman}, W.~{Song}, and A.~{Strominger}, ``{The Kerr/CFT
  correspondence},'' \href{http://dx.doi.org/10.1103/PhysRevD.80.124008}{{\em
  \prd} {\bfseries 80} no.~12, (Dec., 2009) 124008},
  \href{http://arxiv.org/abs/0809.4266}{{\ttfamily arXiv:0809.4266 [hep-th]}}.

\bibitem{Chen:2010qm}
B.~Chen and B.~Ning, ``{Self-Dual Warped AdS$_3$ Black Holes},''
  \href{http://dx.doi.org/10.1103/PhysRevD.82.124027}{{\em Phys. Rev. D}
  {\bfseries 82} (2010) 124027},
  \href{http://arxiv.org/abs/1005.4175}{{\ttfamily arXiv:1005.4175 [hep-th]}}.

\bibitem{Li:2010sv}
R.~Li and J.-R. Ren, ``{Quasinormal Modes of Self-Dual Warped AdS$_3$ Black
  Hole in Topological Massive Gravity},''
  \href{http://dx.doi.org/10.1103/PhysRevD.83.064024}{{\em Phys. Rev. D}
  {\bfseries 83} (2011) 064024},
  \href{http://arxiv.org/abs/1008.3239}{{\ttfamily arXiv:1008.3239 [hep-th]}}.

\end{thebibliography}\endgroup
\bibliographystyle{utphys}

\end{document}